\documentclass{iopart}
\usepackage{iopams}
\usepackage{amssymb}
\usepackage{amsfonts,dsfont}
\usepackage{setstack}
\usepackage{graphicx}
\usepackage{bbold}
\usepackage{color}
\usepackage{hyperref}

\newcommand{\str}{{\rm str\,}}
\newcommand{\sdet}{{\rm sdet\,}}
\newcommand{\diag}{{\rm diag\,}}
\newcommand{\U}{{\rm U\,}}
\newcommand{\UOSp}{{\rm UOSp\,}}
\newcommand{\Gl}{{\rm Gl\,}}
\newcommand{\Herm}{{\rm Herm\,}}
\newcommand{\eins}{{\mathds{1}}}

\eqnobysec

\begin{document}
\title[Supersymmetry for Chiral Random Matrix Theory]{Supersymmetry
  Method for Chiral Random Matrix Theory with Arbitrary Rotation
  Invariant Weights}

\author{Vural Kaymak$\,^{1}$, Mario Kieburg$\,^{2}$, Thomas Guhr$\,^{1}$}

\address{$\,^{1}$ Fakult\"at f\"ur Physik, Universit\"at Duisburg-Essen, Lotharstra\ss e 1, 47048 Duisburg, Germany\\
$\,^{2}$ Fakult\"at f\"ur Physik, Universit\"at Bielefeld, Postfach 100131, 33501 Bielefeld, Germany}

\ead{\mailto{vural.kaymak@uni-due.de}, \mailto{mkieburg@physik.uni-bielefeld.de}, \mailto{thomas.guhr@uni-due.de}}

\begin{abstract}
  In the past few years, the supersymmetry method was generalized to
  real-symmetric, Hermitean, and Hermitean self-dual random matrices
  drawn from ensembles invariant under the orthogonal, unitary, and
  unitary symplectic group, respectively. We extend this supersymmetry
  approach to chiral random matrix theory invariant under the three
  chiral unitary groups in a unifying way. Thereby we generalize a
  projection formula providing a direct link and, hence, a `short cut'
  between the probability density in ordinary space and the one in
  superspace. We emphasize that this point was one of the main
  problems and critiques of the supersymmetry method since only
  implicit dualities between ordinary and superspace were known
  before. As examples we apply this approach to the calculation of the
  supersymmetric analogue of a Lorentzian (Cauchy) ensemble and an
  ensemble with a quartic potential. Moreover we consider the
  partially quenched partition function of the three chiral Gaussian
  ensembles corresponding to four-dimensional continuum QCD. We
  identify a natural splitting of the chiral Lagrangian in its lowest
  order into a part of the physical mesons and a part associated to
  source terms generating the observables, e.g. the level density of
  the Dirac operator.\\

  \noindent{\it Random Matrix Theory, Supersymmetry, Multivariate
    Statistics, Correlated Wishart Matrices, Universality, chiral
    Lagrangian, Multicritical Ensembles, generalized
    Hubbard-Stratonovich transformation, superbosonization formula}
\end{abstract}


\submitto{\JPA}

\maketitle
\section{Introduction}
\label{sec:intro}

Chiral random matrix theory is the oldest of all random matrix
ensembles. It was introduced by Wishart \cite{Wishart} in the 1920's
to model generic properties of correlation matrices. Since then chiral
random matrix theory was applied to many other fields of physics and
beyond because of its versatility. One important application is the
study of correlation matrices in time series analysis
\cite{timeseries1,timeseries2,timeseries3,timeseries4,finance1,finance2,Recher1,Recher2}. Chiral
random matrix theory serves as a benchmark model for empirical
correlation matrices and is used to extract the system specific
correlations from the generic statistical fluctuations. Another famous
development is the introduction of chiral random matrix theory to QCD
by Shuryak and Verbaarschot \cite{RMTQCD1,RMTQCD2,RMTQCDbook}. They
showed the equivalence of the microscopic limit of the QCD-Dirac
operator with chiral random matrix theory. In particular chiral random
matrix theory explained the statistical fluctuations of the smallest
eigenvalues of the Dirac operator and predicted relations between low
energy constants and observables which are confirmed by lattice QCD
data \cite{RMTQCDdata1,RMTQCDdata2}. Recent applications of chiral
random matrix theory can be also found in condensed matter
theory~\cite{conmat1},
telecommunication~\cite{telecom1,telecom2,telecom3}, and quantum
information theory~\cite{inftheo1} but its range is by far not
restricted to those examples.

For the sake of simplicity, a Gaussian function is often used within
the context of random matrix theory. Due to
universality~\cite{univloc1,univloc2,univloc3,univloc4}, this choice
is quite often legitimized as long as the interest lies in
correlations on the local scale of the mean level spacing. To prove
universality as well as to modify random matrix theory to describe
particular systems many technical tool were developed. For example,
the supersymmetry method, originally introduced for Gaussian
weights~\cite{SUSYGauss1,SUSYGauss2,SUSYGauss3,SUSYGauss4}, is
established as a versatile tool in the field of random matrix theory
because of its broad applicability to non-Gaussian ensembles. For the
history of the supersymmetry method and its variants, we refer the
reader to Ref.~\cite{SUSYGauss4}. Moreover, one is not always
interested in the local scale, e.g. see the analysis of universality
on macroscopic scales as it is discussed with free
probability~\cite{univglob1,univglob2}. Insofar a generalization to
arbitrary statistical weights is of particular interest. Other
important techniques are the orthogonal polynomial
method~\cite{OrthogonalPol}, Toda lattice structures~\cite{toda}, free
probability theory~\cite{isotropy} and maps to Hamiltonian
systems~\cite{mapHamilton, mapHamilton2}. For a comprehensive overview
see~\cite{RMTbook,Cauchyens1,Mehta} and references therein.

Here, we focus on the supersymmetry method, not on aspects related to
other methods such as orthogonal polynomials. We start from a close
connection between matrix invariants in ordinary and superspace which
was first observed in Ref.~\cite{genHub1}. In particular for chiral
random matrix models we investigate how probability densities which
only depend on matrix invariants (but are otherwise arbitrary) are
uniquely mapped from ordinary to superspace.  This is the issue at
stake.

An exact map from ordinary space to superspace for arbitrary isotropic
ensembles for real symmetric, Hermitean and Hermitean self-dual
matrices was provided in two different but related approaches, a few
years ago. Isotropy is the invariance under the orthogonal, unitary or
unitary symplectic group, respectively, see Ref.~\cite{isotropy}. One
approach pursues the idea to generalize the original
Hubbard-Stratonovich transformation in superspace for Gaussian
weights~\cite{SUSYGauss1,SUSYGauss2,SUSYGauss3,SUSYGauss4} to
arbitrary weights~\cite{genHub1,genHub2}. In another approach one
tries to find a direct, exact identity between integrals over dyadic
supermatrices and integrals over cosets. This second approach is known
as the superbosonization formula~\cite{supform1,supform2}. Both
approaches are completely equivalent \cite{comp} and both have their
advantages as well as disadvantages. One crucial disadvantage they
both share is that they do not directly relate the probability density
in ordinary space with the one in superspace. They only become
explicit when the characteristic function (Fourier transform of the
probability density) is known in a closed form. Hence one has to
calculate the statistical weight for each random matrix ensemble,
separately.  This is exactly the problem we want to address.

The extension of the generalized Hubbard-Stratonovich transformation
as well as the superbosonization formula to the other seven classes in
the tenfold classification via the Cartan
scheme~\cite{Cartan1,Cartan2} is still unsolved. We address three of
these seven classes in a unifying way, namely chiral random matrices
generated by non-Gaussian probability densities. In particular we
derive a projection formula explicitly relating the probability
density in ordinary space with the one in superspace. Thus we present
a solution to the disadvantage of the generalized Hubbard-Stratonovich
transformation and the superbosonization formula where one has to
study each ensemble separately. Such a projection formula was already
accomplished for real symmetric, Hermitean, and Hermitean self-dual
matrices, see Ref.~\cite{Doktor}. In Sec.~\ref{sec:posingTheProblem},
we briefly summarize the idea behind such a projection formula for
ensembles in the original classification by Dyson~\cite{Cartan1} and
put it into contrast with the well established generalized
Hubbard-Stratonovich transformation and the superbosonization
formula. In Sec.~\ref{sec:Zwei}, we generalize this approach to the
three chiral random matrix theories of real, complex and quaternion
rectangular matrices in a unifying way.

To underline that the projection formula is a powerful tool we apply
it to a selection of ensembles encountered in different fields of
random matrix theory, in Sec.~\ref{sec:examples}. Some of these
ensembles, as the Lorentz (Cauchy)-like ensembles and the ensemble
with a quartic potential, are not at all trivial and it is not
immediately clear what their supersymmetric counterpart will look
like. The other examples are the norm-dependent ensembles without and
with empirical correlations and the unquenched chiral Gaussian random
matrix ensembles modelling QCD with quarks. In particular for the
partially quenched partition function we derive a representation whose
microscopic limit agrees with QCD and shows a natural splitting into
physical mesons and those corresponding to the source term generating
the observables like the level density or higher order
correlations. The explicit calculation of this result is presented in
\ref{app1}. The article is concluded with a summary in
Sec.~\ref{sec:summary}.

\section{Main idea of a projection formula}
\label{sec:posingTheProblem}

The supersymmetry method is essentially a general relation between partition functions in ordinary space,
\begin{eqnarray}\label{partord}
 Z(\kappa)&=&\int d[H] P(H)\frac{\prod\limits_{j=1}^{k_2}\det(H-\kappa_j^{(2)}\eins_N)}{\prod\limits_{j=1}^{k_1}\det(H-\kappa_j^{(1)}\eins_N)},
\end{eqnarray}
and partition functions in superspace, which we expect to be of the form
\begin{eqnarray}\label{partsuper}
 Z(\kappa)&=&\int d[\sigma]Q(\sigma)\sdet^{\mu(N)}(\sigma-\kappa).
\end{eqnarray}
The $N\times N$ matrix $H$ is distributed by $P$ and drawn from one of the Hermitean ensembles classified in the ten-fold way via the Cartan classification scheme \cite{Cartan1,Cartan2}. The exponent $\mu$ is some affine linear function in the former ordinary dimension $N$. The supermatrix $\sigma$ has a dimension related to the number of determinants in Eq.~\eref{partord}. It fulfills certain symmetries depending on the ones of the ordinary matrix $H$, and is drawn from a probability density $Q$ in superspace. The source variables
\begin{equation}
\kappa := \left\{\begin{array}{cl} \mathrm{diag} (\kappa_1, \kappa_2), & \beta = 2,\\ \mathrm{diag} (\kappa_1, \kappa_1, \kappa_2, \kappa_2), & \beta = 1,4,\end{array}\right.
\label{eq:kappaDef2}
\end{equation}
with
\begin{equation}
\kappa_1 := \mathrm{diag}(\kappa_{1}^{(1)}, \ldots, \kappa_{k_1}^{(1)}),\qquad
\kappa_2 := \mathrm{diag}(\kappa_{1}^{(2)}, \ldots, \kappa_{k_2}^{(2)}),
\label{eq:kappaDef}
\end{equation}
are distinguished by the Dyson index $\beta=1,2,4$. We notice that $\kappa$ is always a supermatrix. In the context of QCD, it comprises masses of the physical fermions as well as masses of the valence fermions usually denoted by $m_j$ \cite{RMTQCDbook}. The masses of the valence fermions consist of source variables for differentiation to generate the matrix Green functions often denoted by $J_j$ and markers for the eigenvalues of $H$ which are usually denoted by $x_j$ \cite{SUSYGauss4}. Additionally we have to assume that $\kappa_j^{(1)}$ has a non-zero imaginary part, since the spectrum of $H$ lies on the real axis.

The main task is to derive two things. First of all, the corresponding supermatrix space, $\sigma\in\mathcal{M}_{\rm SUSY}$, has to be identified which is independent of the probability density $P$. This identification was already done in Ref.~\cite{Cartan2}. Second, one has to calculate the probability distribution $Q$ which crucially depends on the ordinary matrix space, $H\in\mathcal{M}_{\rm ord}$, and on the probability density $P$. Exactly the second task is the hardest one and is up to now only known in a closed form when $H$ is real symmetric, Hermitean, or Hermitean self-dual \cite{Doktor}.

After recalling the standard supersymmetry method in subsection~\ref{sec:standard}, we briefly rederive a projection formula for ensembles of real symmetric, Hermitean, and Hermitean self-dual  matrices in  subsection~\ref{sec:brief} to point out the main idea of such a projection formula.

\subsection{Standard supersymmetry approach}\label{sec:standard}

Let us introduce three abbreviations,
\begin{eqnarray}
\fl\mathrm{U}^{(\beta)}(n) &:= &\left\{\begin{array}{cl} \mathrm{O}(n), & \beta=1, \\ \mathrm{U}(n), & \beta=2, \\ \mathrm{USp}(2n), & \beta=4,\end{array}\right.
\label{eq:InvarianzGruppe}\\
\fl\mathrm{Herm}^{(\beta)}(n) &:=& \left\{\begin{array}{cl} {\rm Gl}(n,\mathbb{R})/\mathrm{O}(n)\cong{\rm U}(n)/{\rm O}(n), & \beta=1, \\ {\rm Gl}(n,\mathbb{C})/\mathrm{U}(n), & \beta=2, \\ {\rm Gl}(n,\mathbb{H})/\mathrm{USp}(2n)\cong{\rm U}(2n)/{\rm USp}(2n), & \beta=4,\end{array}\right.
\label{cosetdef}
\end{eqnarray}
and
\begin{eqnarray}\label{gammadef}
 \gamma:=\left\{\begin{array}{cl} 1, &  \beta=1,2, \\ 2, & \beta=4,\end{array}\right.\ {\rm and}\ \widetilde{\gamma}:=\left\{\begin{array}{cl} 2, &  \beta=1, \\ 1, & \beta=2,4,\end{array}\right.
\end{eqnarray}
such that we can deal with all three Dyson indices $\beta=1,2,4$ in a unifying way. Equation~\eref{cosetdef} is an abbreviation for the set of real symmetric, Hermitean, and Hermitean self-dual matrices, respectively. Here, $\mathbb{H}$ is the quaternion number field which we represent via the Pauli matrices and the two-dimensional unit matrix $\eins_2$ throughout the work.

The aim is to identify a partition function in superspace starting from a partition function in ordinary space,
\begin{eqnarray}\label{partord2}
  Z(\kappa)&:=&\int d[H] P(H)\frac{\prod\limits_{j=1}^{k_2}\det(H-\kappa_j^{(2)}\eins_{\gamma n})}{\prod\limits_{j=1}^{k_1}\det(H-\kappa_j^{(1)}\eins_{\gamma n})}\\
 &=&\int d[H] P(H)\sdet^{-1/(\gamma\widetilde{\gamma})}(H\otimes\eins_{\gamma\widetilde{\gamma}k_1|\gamma\widetilde{\gamma}k_2}-\eins_{\gamma n}\otimes\kappa),\nonumber
\end{eqnarray}
with $H\in\mathrm{Herm}^{(\beta)}(n)$ and $P$ fulfilling the rotation invariance (also known as isotropy \cite{isotropy})
\begin{eqnarray}\label{isotropy}
P(H)=P(UHU^{-1}),\quad \forall\ U\in \mathrm{U}^{(\beta)}(n).
\end{eqnarray}
Let for simplicity ${\rm Im}\,\kappa_1>0$ in this subsection. We will weaken this condition later on.

In the original supersymmetry method one introduces a rectangular complex supermatrix $V$ ~\cite{SUSYGauss4} of dimension $(\gamma n)\times (\gamma\widetilde{\gamma}k_1|\gamma\widetilde{\gamma}k_2)$ and uses the crucial identity
\begin{eqnarray}
 \fl\sdet^{-1/(\gamma\widetilde{\gamma})}(H\otimes\eins_{\gamma\widetilde{\gamma}k_1|\gamma\widetilde{\gamma}k_2}-\eins_{\gamma n}\otimes\kappa)=\frac{\int d[V]\exp[\imath\str V^\dagger V\kappa-\imath\str V^\dagger HV]}{\imath^{\gamma n(k_2-k_1)}\int d[V]\exp[-\str V^\dagger V]}.\label{Gaussian}
\end{eqnarray}
Recall the definition of $\kappa$ in Eq.~\eref{eq:kappaDef2} and of $\gamma$ and $\widetilde{\gamma}$ in Eq.~\eref{gammadef}. The rescaling by the imaginary unit $\imath$ is needed to ensure the convergence of the integral over $V$. The supermatrix $V$ consists of independent complex random variables as well as complex Grassmann variables (anti-commuting variables) and fulfills some  symmetries under complex conjugation if the Dyson index is $\beta=1,4$, i.e. the complex conjugate of $V$ is
\begin{eqnarray}\label{symmetries}
 V^*=\left\{\begin{array}{cl} V\diag(\eins_{2k_1},\imath\tau_2\otimes\eins_{k_2}), & \beta=1,\\ (-\imath\tau_2\otimes\eins_n) V\diag(\imath\tau_2\otimes\eins_{k_1},\eins_{2k_2}), & \beta=4, \end{array}\right.
\end{eqnarray}
where $\tau_2$ is the second Pauli matrix. The case $\beta=1$ is some kind of reality condition and for $\beta=4$ it is some kind of generalization of quaternions.

When plugging Eq.~\eref{Gaussian} into  the partition function~\eref{partord2} the integration over $H$ reduces to a  Fourier transform of the probability density $P$.  We assume that the  Fourier transform,
\begin{eqnarray}\label{Fourier}
 \Phi(A):=\int d[H] P(H)\exp[-\imath\tr HA],
\end{eqnarray}
 exists for any $(\gamma n)\times(\gamma n)$ matrix $A$ sharing the same symmetries as $H$ apart from relations involving complex conjugations. The invariance property~\eref{isotropy} of $P$ carries over to one of $\Phi$, i.e.
\begin{eqnarray}\label{isotropyPhi}
\Phi(A)=\Phi(UAU^{-1})\quad \forall\ U\in \mathrm{U}^{(\beta)}(n)
\end{eqnarray}
meaning that the function $\Phi$ can be written as a function of the traces of $A$. Identifying the matrix $A=VV^\dagger$, one can show that there is a superfunction $\widetilde{\Phi}$, which is by far not unique (see Ref.~\cite{genHub2}), such that another essential identity of the supersymmetry method holds \cite{SUSYGauss4},
\begin{eqnarray}\label{duality}
\Phi(VV^\dagger)=\widetilde{\Phi}(V^\dagger V).
\end{eqnarray}
Note that the tilde  emphasizes that $\widetilde{\Phi}$ is not the same as but related to the function $\Phi$. The partition function reads
\begin{eqnarray}\label{partition1}
  Z(\kappa)&=&\frac{\int d[V]\exp[\imath\str V^\dagger V\kappa]\widetilde{\Phi}(V^\dagger V)}{\imath^{\gamma n(k_2-k_1)}\int d[V]\exp[-\str V^\dagger V]}
\end{eqnarray}
which is already a representation in superspace.

Two different ways can be pursued from this point. One approach is the superbosonization formula~\cite{supform1,supform2}. With help of the superbosonization formula the integral over $V^\dagger V$ is replaced by an integral over a $(\gamma\widetilde{\gamma}k_1|\gamma\widetilde{\gamma}k_2)\times(\gamma\widetilde{\gamma}k_1|\gamma\widetilde{\gamma}k_2)$ supermatrix $U$ fulfilling some symmetries under the transposition if $\beta=1,4$, i.e.
\begin{eqnarray}\label{symmetriesU}
 \fl U^T=\left\{\begin{array}{cl} \diag(\eins_{2k_1},-\imath\tau_2\otimes\eins_{k_2})U\diag(\eins_{2k_1},\imath\tau_2\otimes\eins_{k_2}), & \beta=1,\\ \diag(-\imath\tau_2\otimes\eins_{k_1},\eins_{2k_2})U\diag(\imath\tau_2\otimes\eins_{k_1},\eins_{2k_2}), & \beta=4, \end{array}\right.
\end{eqnarray}
which means that $U_{\rm BB}$ is symmetric (self-dual) and $U_{\rm FF}$ is self-dual (symmetric) for $\beta=1$ ($\beta=4$). Additionally, the matrix $U$ consists of four blocks,
\begin{eqnarray}\label{blockrep}
 U=\left[\begin{array}{cc} U_{\rm BB} & \eta^\dagger \\ \eta & U_{\rm FF} \end{array}\right],
\end{eqnarray}
whose off-diagonal blocks $\eta$ and $\eta^\dagger$ contain independent Grassmann variables apart from the condition~\eref{symmetriesU}, the boson-boson block is positive definite, $U_{\rm BB}>0$, and Hermitean, $U_{\rm BB}^\dagger=U_{\rm BB}$, and the fermion-fermion block is unitary, $U_{\rm FF}^\dagger=U_{\rm FF}^{-1}$. Hence the supermatrix $U$ is in one of the three cosets~\cite{supform1,supform2,comp}
\begin{eqnarray}\label{superHerm}
 \fl\Herm_{\odot}^{(\beta)}(\widetilde{\gamma}k_1|\gamma k_2):=\left\{\begin{array}{cl} \U(2k_1|2k_2)/\UOSp^{(+)}(2k_1|2k_2), & \beta=1,\\ \Gl(k_1|k_2)/\U(k_1|k_2), & \beta=2,\\ \U(2k_1|2k_2)/\UOSp^{(-)}(2k_1|2k_2), & \beta=4,\end{array}\right.
\end{eqnarray}
where $\U(p|q)$ is the unitary supergroup and $\Gl(p|q)$ is the general linear, complex supergroup.
The two supergroups $\UOSp^{(\pm)}(2k_1|2k_2)$ for $\beta=1,4$ are the two independent matrix-representations of the unitary ortho-symplectic supergroup $\UOSp(2k_1|2k_2)$. Matrices in this group are real in the boson-boson block and quaternion in the fermion-fermion block for $\beta=1$ denoted by the superscript ``$(+)$'' and vice versa for $\beta=4$ denoted by the superscript ``$(-)$'', see Ref.~\cite{genHub2}. The subscript ``$\odot$'' refers to the kind of embedding of the coset which is a contour-integral around the origin for the fermion-fermion block $U_{\rm FF}$ in the case of the superbosonization formula.

The superbosonization formula can be summarized to  the following simple equation,
\begin{eqnarray}\label{partition2}
  Z(\kappa)&=&\frac{\int d\mu(U)\sdet^{n/\widetilde{\gamma}}U\exp[\imath\str U\kappa]\widetilde{\Phi}(U)}{\imath^{\gamma n(k_2-k_1)}\int d\mu(U) \sdet^{n/\widetilde{\gamma}}U\exp[-\str U]},
\end{eqnarray}
see Refs.~\cite{supform1,supform2}. The measure $d\mu(U)$ is the Haar measure of the corresponding coset.

The second supersymmetric approach is the generalized Hubbard-Stratonovich transformation~\cite{genHub1,genHub2}. Instead of replacing $V^\dagger V$ by a supermatrix one assumes that the superfunction $\widetilde{\Phi}$ is a Fourier transform of another superfunction $Q$ as well, i.e.
\begin{eqnarray}\label{Fourierback}
 \widetilde{\Phi}(B)=\int d[\sigma] Q(B)\exp[-\imath\str \sigma B],
\end{eqnarray}
for some supermatrix $B$. The integration domain of $\sigma$ is very important. First of all it fulfills the same symmetries under transposition as $U$ in the superbosonization formula, see Eq.~\eref{symmetriesU}, i.e.
\begin{eqnarray}\label{symmetriessig}
 \sigma^T=\left\{\begin{array}{cl} \diag(\eins_{2k_1},-\imath\tau_2\otimes\eins_{k_2})\sigma\diag(\eins_{2k_1},\imath\tau_2\otimes\eins_{k_2}), & \beta=1,\\ \diag(-\imath\tau_2\otimes\eins_{k_1},\eins_{2k_2})\sigma\diag(\imath\tau_2\otimes\eins_{k_1},\eins_{2k_2}), & \beta=4, \end{array}\right.
\end{eqnarray}
which is again equivalent that $\sigma_{\rm BB}$ is symmetric (self-dual) and $\sigma_{\rm FF}$ is self-dual (symmetric) for $\beta=1$ ($\beta=4$). 
However the blocks of
\begin{eqnarray}\label{blockrepsig}
 \sigma=\left[\begin{array}{cc} \sigma_{\rm BB} & \eta^\dagger \\ \eta & \sigma_{\rm FF} \end{array}\right]
\end{eqnarray}
are drawn from different supports as for $U$. The off-diagonal blocks $\eta$ and $\eta^\dagger$ are again independent Grassmann variables apart from the condition~\eref{symmetriessig} while the boson-boson block is now only Hermitean, $\sigma_{\rm BB}^\dagger=\sigma_{\rm BB}$. The fermion-fermion block can be diagonalized by $\widehat{U}\in\U^{(4/\beta)}(\gamma k_2)$, i.e. $\sigma_{\rm FF}=\widehat{U}s_{\rm FF}\widehat{U}^\dagger$. The eigenvalues $\{s_{\rm FF}\}_j$ live on contours such that the integral over them converges. For a Gaussian ensemble the standard Wick-rotation, i.e. $\{s_{\rm FF}\}_j\in \imath\mathbb{R}$, does the job. For other polynomial potentials one has to choose other Wick-rotations, e.g. for $P(H)\propto\exp[-\tr H^{2m}]$ it is $\{s_{\rm FF}\}_j\in e^{\imath\pi/(2m)}\mathbb{R}$. Therefore the supermatrix $\sigma$ lies also in an embedding of the cosets~\eref{superHerm} but the set will be now denoted by $\Herm_{\rm Wick}^{(\beta)}(\widetilde{\gamma}k_1
 |\gamma k_2)$ where the subscript ``${\rm Wick}$'' reflects the nature of the integration domain.

Reading off $B=V^\dagger V$ and integrating over $V$ one obtains the final result for the generalized Hubbard-Stratonovich transformation,
\begin{eqnarray}\label{partition3}
  Z(\kappa)&=&\int d[\sigma]Q(\sigma)\sdet^{-n/\widetilde{\gamma}}(\sigma-\kappa),
\end{eqnarray}
see Refs.~\cite{genHub1,genHub2}. The measure $d[\sigma]$ is the flat one, i.e. the product of the differential of all independent matrix elements.

Both approaches, the superbosonization formula as well as the generalized Hubbard-Stratonovich transformation, have a crucial weakness. Without an explicit knowledge of the Fourier transform $\Phi$ no direct functional relation between the probability density $P$, the superfunction $\widetilde{\Phi}$, and the superfunction $Q$ is known. The reason is the duality relation~\eref{isotropyPhi} between ordinary and superspace. Particularly for the generalized Hubbard-Stratonovich transformation, the dyadic matrices $VV^\dagger$ and $V^\dagger V$ are in different matrix spaces. Hence, one cannot expect that the Fourier transforms~\eref{Fourier} and \eref{Fourierback} yield the same functional dependence of $P$ and $Q$. The projection formula~\cite{Doktor} briefly rederived in subsection~\ref{sec:brief} circumvents this problem.

\subsection{Projection formula for Dyson's threefold way}\label{sec:brief}

The key idea to find a direct relation between $P$ and $Q$ is to extend the original matrix set $H\in\mathrm{Herm}^{(\beta)}(n)$ to a larger matrix set also comprising the target set $\sigma\in\Herm_{\rm Wick}^{(\beta)}(\widetilde{\gamma}k_1|\gamma k_2)$. Let $\widetilde{\gamma}k_1\leq \widetilde{\gamma}k_2+n$ to keep the calculation as simple as possible otherwise we have to do a case discussion. This condition is usually the case when applying supersymmetry to random matrix theory. Nevertheless we underline that this condition is not at all a restriction since the other case can be taken care by slightly modifying the ensuing discussion, see Ref.~\cite{Doktor}.

The idea of our approach is based on a Cauchy-like integration formula for supermatrices in the coset $\Herm_{\rm Wick}^{(\beta)}(p|q)$ with $p,q\in\mathbb{N}_0$ which was first derived by Wegner~\cite{Cauchy1}, see also Refs.~\cite{Cauchy2,Cauchy3,Cauchy4} for slightly modified versions. Let $l\in\mathbb{N}$ be a positive integer and $f$ be an integrable and smooth superfunction on the set of supermatrices $\Herm_{\rm Wick}^{(\beta)}(p+\widetilde{\gamma}l|q+\gamma l)$ and invariant under
\begin{eqnarray}\label{invariance}
 f(\widetilde{U}\Sigma\widetilde{U}^{-1})=f(\Sigma)
\end{eqnarray}
for all $\Sigma\in\Herm_{\rm Wick}^{(\beta)}(p+\widetilde{\gamma}l|q+\gamma l)$ and
\begin{eqnarray}\label{superuni}
  \fl\widetilde{U}\in \U^{(\beta)}(p+\widetilde{\gamma}l|q+\gamma l):=\left\{\begin{array}{cl} \UOSp^{(+)}(p+2l|2q+2l), & \beta=1,\\ \U(p+l|q+l), & \beta=2,\\ \UOSp^{(-)}(2p+2l|q+2l), & \beta=4.\end{array}\right.
\end{eqnarray}
Employing the following splitting of
\begin{eqnarray}\label{splitting}
 \Sigma=\left[\begin{array}{cc} \widetilde{\Sigma} &  0 \\ 0 & 0 \end{array}\right]+\widehat{\Sigma}\quad {\rm with}\quad \widehat{\Sigma}=\left[\begin{array}{cc} 0 & \widehat{V} \\ \widehat{V}^\dagger & \sigma \end{array}\right]
\end{eqnarray}
such that $\widetilde{\Sigma}\in\Herm_{\rm Wick}^{(\beta)}(p|q)$ and $\sigma\in\Herm_{\rm Wick}^{(\beta)}(\widetilde{\gamma}l|\gamma l)$, the Cauchy-like integral identity~\cite{Cauchy1, Cauchy2, Cauchy3, Cauchy4} reads
\begin{eqnarray}\label{Cauchy-Hermite}
 \frac{\int d[\widehat{\Sigma}] f(\Sigma)}{\int d[\widehat{\Sigma}]\exp[-\str\widehat{\Sigma}^2]}=f\left(\left[\begin{array}{cc} \widetilde{\Sigma} &  0 \\ 0 & 0 \end{array}\right]\right)
\end{eqnarray}
reducing a large supermatrix, $\Sigma$, to a smaller one, $\widetilde{\Sigma}$, independent of the concrete form of the superfunction $f$. The notation $\Sigma$, $\widetilde{\Sigma}$ and $\widehat{\Sigma}$ has no deeper meaning. It only underlines that all three matrices are essentially of the same form apart from their different dimensions.

Equation~\eref{Cauchy-Hermite} is at the heart of our approach. Let us consider the partition function~\eref{partord2} in ordinary space. From now on, we lift the condition ${\rm Im}\,\kappa_1>0$ to emphasize that our idea works in general and define $L={\rm sign}\,{\rm Im}\,\kappa$. We assume that the probability density $P$ is rotation invariant, see Eq.~\eref{isotropy}. Moreover we assume that a contour like the Wick-rotation and an extension of $P$, denoted by $\widetilde{P}$, from the ordinary matrix set $\Herm^{(\beta)}(n)$ to the supermatrix set  $\Herm_{\rm Wick}^{(\beta)}(n+\widetilde{\gamma} k_2|\gamma k_2)$ exists such that the superfunction $\widetilde{P}$ is integrable and smooth on $\Herm_{\rm Wick}^{(\beta)}(n+\widetilde{\gamma} k_2|\gamma k_2)$. Then we can extend the integral~\eref{partord2} to an integral in superspace, i.e.
\begin{eqnarray}\label{partpro1}
  Z(\kappa) &=&\frac{\int d[\Sigma] \widetilde{P}(\Sigma)\sdet^{-1/(\gamma\widetilde{\gamma})}(H\otimes\eins_{\gamma\widetilde{\gamma}k_1|\gamma\widetilde{\gamma}k_2}-\eins_{\gamma n}\otimes\kappa)}{\int d[\widehat{\Sigma}]\exp[-\str\widehat{\Sigma}^2]},
\end{eqnarray}
where we employ a splitting similar to Eq.~\eref{splitting}, i.e.
\begin{eqnarray}
 \fl\Sigma=\left[\begin{array}{cc} H &  0 \\ 0 & 0 \end{array}\right]+\widehat{\Sigma}=\left[\begin{array}{cc} 0 &  0 \\ 0 & \sigma \end{array}\right]+\Sigma'\  {\rm with}\  \widehat{\Sigma}=\left[\begin{array}{cc} 0 & \widehat{V} \\ \widehat{V}^\dagger & \widehat{\sigma} \end{array}\right]\ {\rm and}\ \Sigma'=\left[\begin{array}{cc} H' & V' \\ V^{\prime\,\dagger} & 0 \end{array}\right]\nonumber\\
 \label{splittingsig}
\end{eqnarray}
 with $H\in\Herm_{\rm Wick}^{(\beta)}(n|0)=\Herm^{(\beta)}(n)$, $H'\in\Herm_{\rm Wick}^{(\beta)}(n+\widetilde{\gamma}(k_2-k_1)|0)=\Herm^{(\beta)}(n+\widetilde{\gamma}(k_2-k_1))$, $\widehat{\sigma}\in\Herm_{\rm Wick}^{(\beta)}(\widetilde{\gamma}k_2|\gamma k_2)$, and $\sigma\in\Herm_{\rm Wick}^{(\beta)}(\widetilde{\gamma}k_1|\gamma k_2)$.
 The second splitting becomes more important later on. Notice that we extended $H\to\Sigma$ in the probability density $\widetilde{P}$, only.

From now on we pursue the ideas of the standard supersymmetry method, see subsection~\ref{sec:standard}. We introduce the same rectangular supermatrix $V$ as in Eq.~\eref{Gaussian}, i.e.
\begin{eqnarray}
 \fl\sdet^{-1/(\gamma\widetilde{\gamma})}(H\otimes\eins_{\gamma\widetilde{\gamma}k_1|\gamma\widetilde{\gamma}k_2}-\eins_{\gamma n}\otimes\kappa)=\frac{\int d[V]\exp[\imath\str V^\dagger V L\kappa-\imath\str V^\dagger HVL]}{\imath^{\gamma n(k_2-k_1)}\sdet^{-n/\widetilde{\gamma}} L\int d[V]\exp[-\str V^\dagger V]}.\nonumber\\
 \fl\label{Gaussian2}
\end{eqnarray}
In terms of $\Sigma$ the partition function reads
\begin{eqnarray}\label{partpro2}
  \fl Z(\kappa) &=&\frac{\int d[\Sigma] \widetilde{P}(\Sigma)\int d[V]\exp[\imath\str V^\dagger V L\kappa-\imath\str \Sigma \widehat{A}]}{\imath^{\gamma n(k_2-k_1)}\sdet^{-n/\widetilde{\gamma}} L\int d[V]\exp[-\str V^\dagger V]\int d[\widehat{\Sigma}]\exp[-\str\widehat{\Sigma}^2]}
  \end{eqnarray}
  with
  \begin{eqnarray}\label{ordsplit}
   \widehat{A}=\left[\begin{array}{cc} VLV^\dagger &  0 \\ 0 & 0 \end{array}\right]=\left[\begin{array}{cc} 0 & V\sqrt{L}\\  0 & 0 \end{array}\right]\left[\begin{array}{cc} 0 & 0 \\ \sqrt{L}V^\dagger &  0 \end{array}\right]
  \end{eqnarray}
  and $\sqrt{L}$ the positive root of the diagonal elements of $L$. The block structure of $\widehat{A}$ corresponds to the first splitting of $\Sigma$ in Eq.~\eref{splittingsig}. The Fourier-Laplace transform
  \begin{eqnarray}\label{Fourier-Laplace}
  \widehat{\Phi}(\widehat{A})=\int d[\Sigma] \widetilde{P}(\Sigma)\exp[-\imath\str \Sigma \widehat{A}]
  \end{eqnarray}
is assumed to exist such that we can interchange the integrals over $\Sigma$ and $V$. Employing the same symmetry arguments as in Eq.~\eref{isotropyPhi} we have
  \begin{eqnarray}\label{duality-new}
  \fl\widehat{\Phi}(\widehat{A})=\widehat{\Phi}(\widehat{B})\quad{\rm with}\quad 
   \widehat{B}=\left[\begin{array}{cc} 0 & 0 \\ \sqrt{L}V^\dagger &  0 \end{array}\right]\left[\begin{array}{cc} 0 & V\sqrt{L} \\  0  & 0 \end{array}\right]=\left[\begin{array}{cc} 0 &  0 \\ 0 & \sqrt{L}V^\dagger V\sqrt{L} \end{array}\right].
  \end{eqnarray}
  The block structure of $\widehat{B}$ is the one of the second splitting of $\Sigma$ in Eq.~\eref{splittingsig}. The advantage of Eq.~\eref{duality-new} in contrast to Eq.~\eref{duality} is that the superfunction $\widehat{\Phi}$ is still the same since $\widehat{A}$ and $\widehat{B}$ are in the same supermatrix set. Hence the inverse Fourier transform is still $\widetilde{P}$ and {\it not} some new superfunction.
  
  The only technical difficulty grows from a non-trivial $L$ because we cannot simply exchange the integrations over $\Sigma$ and $V$ again. To overcome this problem we introduce an auxiliary supermatrix $\sigma_{\rm aux}\in\Herm_{\imath}^{(\beta)}(\widetilde{\gamma}k_1|\gamma k_2)$ drawn from a Gaussian distribution where the subscript ``$\imath$" denotes the standard Wick-rotation~\cite{SUSYGauss1,SUSYGauss2,SUSYGauss3,SUSYGauss4} by the imaginary unit. This Gaussian models some kind of Dirac $\delta$-function, i.e. we can ``simplify''
\begin{eqnarray}
  \fl \exp[-\imath\str \sqrt{L}\sigma\sqrt{L}V^\dagger V]&=&\lim\limits_{t\to0}\frac{\int d[\sigma_{\rm aux}]\exp[-\str(\sigma_{\rm aux}-\sqrt{L}\sigma\sqrt{L})^2/t-\imath\str \sigma_{\rm aux}V^\dagger V]}{\int d[\sigma_{\rm aux}]\exp[-\str\sigma_{\rm aux}^2/t]},\nonumber\\
  \fl\label{Gaussext}
  \end{eqnarray}
  where $t/2$ is the variance of the Gaussian distribution. Assuming that the integral of $\widetilde{P}$ multiplied with $\exp[|\str\sigma^2|]$ exists, we are allowed to interchange the integrations over $\Sigma$, $V$, and $\sigma_{\rm aux}$. We underline that the integrability of $\widetilde{P}$ with $\exp[|\str\sigma^2|]$ is a weak restriction which can be lifted at the end of the day; for example a modification of $P(H)$ to $P(H)\exp[-\delta\tr H^4]$ ($\delta>0$)  does the job and we can take $\delta\to 0$ in the end.
  
  After introducing $\sigma_{\rm aux}$ we interchange the integrals and integrate over $V$ first. Shifting $\sigma_{\rm aux}$ by $\sqrt{L}\sigma\sqrt{L}$ we can take the limit $t\to0$. Finally the partition function takes the simple form
\begin{eqnarray}\label{partpro3}
  Z(\kappa) &=&\frac{\int d[\Sigma] \widetilde{P}(\Sigma)\sdet^{-n/\widetilde{\gamma}}(\sigma-\kappa)}{\int d[\widehat{\Sigma}]\exp[-\str\widehat{\Sigma}^2]}.
  \end{eqnarray}
  Notice that the superdeterminant only depends on $\sigma$ and not anymore on the ordinary matrix $H$.
  
 In the last step we  identify the superfunction $Q$ by comparing the result~\eref{partpro3} with the result of the generalized Hubbard-Stratonovich transformation~\eref{partition3} yielding the final result of this section which is the projection formula
  \begin{eqnarray}\label{projection1}
  Q(\sigma)=\frac{\int d[\Sigma'] \widetilde{P}\left(\left[\begin{array}{cc} 0 &  0 \\ 0 & \sigma \end{array}\right]+\Sigma'\right)}{\int d[\widehat{\Sigma}]\exp[-\str\widehat{\Sigma}^2]}.
  \end{eqnarray}
We integrate over different splittings of $\Sigma$ in the numerator and the denominator. Recall the definition~\eref{splittingsig} of the matrices $\widehat{\Sigma}$ and $\Sigma'$. The superfunction $\widetilde{\Phi}$ in the superbosonization formula~\eref{partition2} can be obtained by the Fourier transformation~\eref{Fourierback} of $Q$.

We underline that the projection formula also holds if the source $\kappa$ is chosen non-diagonal as it sometime happens in QCD \cite{genmass} or if we add an external operator $H_0$ to the original random matrix $H$ often consider in transition ensembles \cite{GUEtrans1,GUEtrans2}. In both cases the integral~\eref{partpro3} is slightly modified but the fundamental functional relation~\eref{projection1} still remains the same.

The projection formula~\eref{projection1} has one big advantage which the results of the superbosonization formula~\eref{partition2} and of the generalized Hubbard-Stratonovich transformation~\eref{partition3} are lacking. With the aid of the projection formula one can study deformations of the probability weight in a quite elegant way. Exactly such an advantage we want to achieve for the chiral ensembles, too.

Finally, we emphasize that the projection formula~\eref{projection1}, after extending $P$ to $\widetilde{P}$, yields one of infinitely many probability weights in superspace corresponding to the same partition function in ordinary space~\eref{partord2}.  This ambiguity of the weight in superspace is well known \cite{genHub2}. Moreover other extensions of $P$ to superspace certainly result into other superfunctions $Q$. Thus an interesting mathematical question is: When varying over all possible extensions $\widetilde{P}$ of $P$, do we get all possible probability weights $Q$ in superspace obtained by the generalized Hubbard-Stratonovich transformation, agreeing with exactly the same partition functions in ordinary space?

\section{Projection formula for chiral ensembles}
\label{sec:Zwei}

The aim is to generalize the projection formula~\eref{projection1} to chiral ensembles. We introduce the chiral matrix
\begin{eqnarray}\label{Hchi}
 H_{\chi}=\left[\begin{array}{cc} 0 & W \\ W^\dagger & 0 \end{array}\right],
\end{eqnarray}
where the matrix entries of $W$ are either real, complex, or quaternion independent random variables for $\beta=1,2,4$, respectively. The chiral matrix $H_{\chi}$ is related to the anti-Hermitean, chiral random matrix
\begin{equation}
\mathcal{D} \longrightarrow D = \left[\begin{array}{cc} 0 & W \\ -W^\dagger & 0 \end{array}\right]=\gamma_5H_{\chi} ,\quad{\rm with}\ \gamma_5=(\eins_n,-\eins_{n+\nu})
\label{eq:DmatrixRepr}
\end{equation}
modelling the Euclidean Dirac operator $\mathcal{D}$ in four dimensions \cite{RMTQCD1,RMTQCD2,RMTQCDbook}. The modulus of the index $\nu\in\{-n,1-n, 2-n,\ldots\}$ is equal to the number of generic zeros of $H_{\chi}$ which can be identified with the topological charge in continuum theory. The random matrix $W$ is drawn from the coset
\begin{eqnarray}\label{Chiens}
 \Gl^{(\beta)}(n;n+\nu):=\U^{(\beta)}(2n+\nu)/[\U^{(\beta)}(n)\times\U^{(\beta)}(n+\nu)]
\end{eqnarray}
 distributed by $P_\chi$ such that $H_\chi\in\Herm^{(\beta)}(2n+\nu)$. The probability density is assumed to be invariant under
\begin{eqnarray}\label{invariancechi}
 P_\chi(W)=P_\chi(UW),\quad\forall\ U\in\U^{(\beta)}(n).
\end{eqnarray}
Notice that we do not assume invariance under right transformations as well which is usually the case \cite{Mehta,RMTQCDbook}. The reason is that we also want to study correlated random matrix ensembles as they naturally appear in the analysis of one-sided correlated Wishart ensembles where the invariance is broken by an empirical correlation matrix, see Refs.~\cite{timeseries1,timeseries2,timeseries3,finance1,finance2,Recher1,Recher2}.

Due to the invariance~\eref{invariancechi} we can reduce the functional dependence of $P_\chi$ on $W$ to one of $WW^\dagger$. Thus there is a function $P$ such that
\begin{eqnarray}\label{reduction}
 P_\chi(W)=P(W^\dagger W).
\end{eqnarray}
Moreover we assume that the chiral partition function,
\begin{eqnarray}\label{partordchi1}
  \fl Z_\chi(\kappa)&:=&\int d[W] P_\chi(H_\chi)\frac{\prod\limits_{j=1}^{k_2}\det(H_\chi-\kappa_j^{(2)}\eins_{\gamma (2n+\nu)})}{\prod\limits_{j=1}^{k_1}\det(H_\chi-\kappa_j^{(1)}\eins_{\gamma (2n+\nu)})}\\
 \fl&=&\int d[W] P_\chi(H_\chi)\sdet^{-1/(\gamma\widetilde{\gamma})}(H_\chi\otimes\eins_{\gamma\widetilde{\gamma}k_1|\gamma\widetilde{\gamma}k_2}-\eins_{\gamma  (2n+\nu)}\otimes\kappa),\nonumber
\end{eqnarray}
can be reduced to one for $W W^\dagger$ or/and $W^\dagger W$,
\begin{eqnarray}
  \fl Z_\chi(\kappa)&=&(-1)^{\gamma (n+\nu)(k_2-k_1)}\sdet^{-\nu/\widetilde{\gamma}}\kappa\label{partordchi2}\\
 \fl&&\times\int d[W] P(W^\dagger W)\sdet^{-1/(\gamma\widetilde{\gamma})}(WW^\dagger\otimes\eins_{\gamma\widetilde{\gamma}k_1|\gamma\widetilde{\gamma}k_2}-\eins_{\gamma  n}\otimes\kappa^2),\nonumber\\
  \fl&=&(-1)^{\gamma (n+\nu)(k_2-k_1)}\sdet^{\nu/\widetilde{\gamma}}\kappa\nonumber\\
 \fl&&\times\int d[W] P(W^\dagger W)\sdet^{-1/(\gamma\widetilde{\gamma})}(W^\dagger W\otimes\eins_{\gamma\widetilde{\gamma}k_1|\gamma\widetilde{\gamma}k_2}-\eins_{\gamma  (n+\nu)}\otimes\kappa^2).\nonumber
\end{eqnarray}
One has to understand that those partition functions do not cover all interesting spectral correlation functions. For example QCD with finite chemical potential or/and finite temperature cannot be modelled with this restriction, cf. Refs.~\cite{finiteQCD1,finiteQCD2,finiteQCD3,RMTQCDbook}. For those partition functions the approach of a projection formula can be modified. Unluckily this modified approach only works for the case $\beta=2$. We will elaborate more on this problem in a forthcoming publication \cite{VKG}.

To make contact with the projection formula~\eref{projection1} for the original ensembles in Dyson's threefold way, we notice that the second representation of the partition function in Eq.~\eref{partordchi2} can be expressed in terms of an integral over $H\in\Herm^{(\beta)}(n)$ if $\nu\leq0$,
\begin{eqnarray}
   Z_\chi(\kappa)&\propto&\sdet^{\nu/\widetilde{\gamma}}\kappa\int d[H] \Theta(H) {\det}^{|\nu|/\widetilde{\gamma}+(\gamma-\widetilde{\gamma})/2} HP(H)\label{partordchi3}\\
 &&\times \sdet^{-1/(\gamma\widetilde{\gamma})}(H\otimes\eins_{\gamma\widetilde{\gamma}k_1|\gamma\widetilde{\gamma}k_2}-\eins_{\gamma ( n+\nu)}\otimes\kappa^2),\nonumber\nonumber
\end{eqnarray}
with the matrix version of the Heaviside $\Theta$ function. It is unity if $H$ is positive definite and otherwise vanishes. Apart from the similarity of Eq.~\eref{partordchi3} with Eq.~\eref{partord2} by identifying $\Theta(H) {\det}^{|\nu|/\widetilde{\gamma}+(\gamma-\widetilde{\gamma})/2} HP(H)$ as the new probability density, the crucial differences are the non-isotropy of $P$, i.e. Eq.~\eref{isotropy} does not necessarily  apply, and the Heaviside $\Theta$ function which is by far not smooth. Thus the original projection formula~\eref{projection1} is not applicable anymore.

In subsection~\ref{sec:projectionchi}, we pursue a similar idea as presented in subsection~\ref{sec:brief} to find a projection formula for partition functions of the form~\eref{partordchi2}. This formula is simplified via a combination with the superbosonization formula in subsection~\ref{sec:projsusy}.

\subsection{Projection formula}
\label{sec:projectionchi}

The key idea to derive a projection formula is again to apply one of the Cauchy-like integration theorems for supermatrices first derived by Wegner~\cite{Cauchy1}, see also Refs.~\cite{Cauchy2,Cauchy3,Cauchy4}. This time we need a Cauchy-like integration theorem for extending the set of rectangular matrices $\Gl^{(\beta)}(n;n+\nu)$ to a space of rectangular supermatrices which is the coset
\begin{eqnarray}\label{Chienssusy}
 \fl \Gl^{(\beta)}(n+\widetilde{\gamma}l|\gamma l;n+\nu):=\U^{(\beta)}(2n+\nu+\widetilde{\gamma}l|\gamma l)/[\U^{(\beta)}(n+\widetilde{\gamma}l|\gamma l)\times\U^{(\beta)}(n+\nu)]
\end{eqnarray}
with $l\in\mathbb{N}$.

Let $p_1,p_2,q,l\in\mathbb{N}_0$. We split a rectangular $(p_1+\widetilde{\gamma}l|q+\gamma l)\times p_2$ supermatrix $\Omega$ in the following way
\begin{eqnarray}\label{Chisplit}
 \Omega=\left[\begin{array}{c} \widetilde{\Omega}  \\ \widehat{\Omega} \end{array}\right]\in\Gl^{(\beta)}(p_1+\widetilde{\gamma}l|q+\gamma l;p_2)
\end{eqnarray}
with $\widetilde{\Omega}\in\Gl^{(\beta)}(p_1|q;p_2)$ and $\widehat{\Omega}\in\Gl^{(\beta)}(\widetilde{\gamma}l|\gamma l;p_2)$. Assuming a smooth superfunction $f$ integrable on the set $\Gl^{(\beta)}(p_1+\widetilde{\gamma}l|q+\gamma l;p_2)$ and invariant under
\begin{eqnarray}\label{invchif}
\fl f(\Omega)=f(\widetilde{U}\Omega),\quad\forall\ \widetilde{U}\in\U^{(\beta)}(p_1+\widetilde{\gamma}l|q+\gamma l)\ {\rm and}\ \Omega\in\Gl^{(\beta)}(p_1+\widetilde{\gamma}l|q+\gamma l;p_2),
\end{eqnarray}
the Cauchy-like integration theorem for rectangular supermatrices \cite{Cauchy1,Cauchy2,Cauchy3,Cauchy4} reads
\begin{eqnarray}\label{Inttheochi}
 \frac{\int d[\widehat{\Omega}] f(\Omega)}{\int d[\widehat{\Omega}] \exp[-\tr\widehat{\Omega}^\dagger\widehat{\Omega}]}=f\left(\left[\begin{array}{c} \widetilde{\Omega} \\ 0 \end{array}\right]\right).
\end{eqnarray}
We notice that no Wick-rotation is needed for this theorem in contrast to Eq.~\eref{Cauchy-Hermite}, simplifying the derivation by getting rid of one technical detail.

We apply the identity~\eref{Inttheochi} to the partition function
\begin{eqnarray}
  \fl Z_\chi(\kappa)&=&(-1)^{\gamma (n+\nu)(k_2-k_1)}\sdet^{-\nu/\widetilde{\gamma}}\kappa\label{partchi1}\\
 \fl&&\times\int d[W] P(W^\dagger W)\sdet^{-1/(\gamma\widetilde{\gamma})}(WW^\dagger \otimes\eins_{\gamma\widetilde{\gamma}k_1|\gamma\widetilde{\gamma}k_2}-\eins_{\gamma  n}\otimes\kappa^2).\nonumber
\end{eqnarray}
We have chosen the first version of Eq.~\eref{partordchi2}, the reason for this choice becomes clearer later on. The product $W^\dagger W$ and, hence, the function $P(W^\dagger W)$ are obviously invariant under left multiplication of $W$ with unitary matrices and can, thus, generally be extended to $\Omega^\dagger\Omega$ and $P(\Omega^\dagger\Omega)$ by the integration theorem~\eref{Inttheochi}, respectively. The only thing we assume is that $P(\Omega^\dagger\Omega)$ has to be smooth and integrable on $\Gl^{(\beta)}(n+\widetilde{\gamma}k_2|\gamma k_2;n+\nu)$ where we again restrict ourself to the case $\widetilde{\gamma}k_1\leq\widetilde{\gamma}k_2+ n$. The other, usually less interesting case $\widetilde{\gamma}k_1\geq\widetilde{\gamma}k_2+ n$ can be derived in a slightly modified discussion.

In the first step we apply the Cauchy-like integration theorem to the partition function to extend the integral over the ordinary space $\Gl^{(\beta)}(n;n+\nu)$ to an integral over the superspace $\Gl^{(\beta)}(n+\widetilde{\gamma}k_2|\gamma k_2;n+\nu)$, i.e.
\begin{eqnarray}
  \fl Z_\chi(\kappa)&=&(-1)^{\gamma (n+\nu)(k_2-k_1)}\sdet^{-\nu/\widetilde{\gamma}}\kappa\label{partchi2}\\
 \fl&&\times\frac{\int d[\Omega] P(\Omega^\dagger\Omega)\sdet^{-1/(\gamma\widetilde{\gamma})}(WW^\dagger \otimes\eins_{\gamma\widetilde{\gamma}k_1|\gamma\widetilde{\gamma}k_2}-\eins_{\gamma  n}\otimes\kappa^2)}{\int d[\widehat{\Omega}] \exp[-\tr\widehat{\Omega}^\dagger\widehat{\Omega}]},\nonumber
\end{eqnarray}
where we employ the following splitting of the rectangular supermatrix,
\begin{eqnarray}\label{Chisplit2}
 \Omega=\left[\begin{array}{c} W  \\ \widehat{\Omega} \end{array}\right]=\left[\begin{array}{c} W'  \\ \Omega' \end{array}\right]
\end{eqnarray}
with $W\in\Gl^{(\beta)}(n|0;n+\nu)=\Gl^{(\beta)}(n;n+\nu)$, $W'\in\Gl^{(\beta)}(n+\widetilde{\gamma}(k_2-k_1)|0;n+\nu)=\Gl^{(\beta)}(n+\widetilde{\gamma}(k_2-k_1);n+\nu)$, $\widehat{\Omega}\in\Gl^{(\beta)}(\widetilde{\gamma}k_2|\gamma k_2;n+\nu)$, and  $\Omega'\in\Gl^{(\beta)}(\widetilde{\gamma}k_1|\gamma k_2;n+\nu)$. The second splitting corresponds to the embedding of the superspace we aim at.

Let $\widetilde{L}={\rm sign}\,{\rm Im}\,\kappa^2$ be the sign of the squared source variables arrayed on a diagonal matrix. In the next step of our approach we introduce Gaussian integrals over exactly the same rectangular supermatrix $V$ as in Eq.~\eref{Gaussian} yielding
\begin{eqnarray}
  \fl Z_\chi(\kappa)&=&(-1)^{\gamma (n+\nu)(k_2-k_1)}\sdet^{-\nu/\widetilde{\gamma}}\kappa\label{partchi3}\\
 \fl&&\times\frac{\int d[\Omega] P(\Omega^\dagger\Omega)\int d[V]\exp[\imath\str V^\dagger V \widetilde{L}\kappa^2-\imath\str \Omega\Omega^\dagger \widetilde{A}]}{\imath^{\gamma n(k_2-k_1)}\sdet^{-n/\widetilde{\gamma}} \widetilde{L}\int d[V]\exp[-\str V^\dagger V]\int d[\widehat{\Omega}] \exp[-\tr\widehat{\Omega}^\dagger\widehat{\Omega}]}\nonumber
\end{eqnarray}
with
  \begin{eqnarray}\label{ordsplit2}
   \widetilde{A}=\left[\begin{array}{cc} V\widetilde{L}V^\dagger &  0 \\ 0 & 0 \end{array}\right]=\left[\begin{array}{cc} 0 & V\sqrt{\widetilde{L}}\\  0 & 0 \end{array}\right]\left[\begin{array}{cc} 0 & 0 \\ \sqrt{\widetilde{L}}V^\dagger &  0 \end{array}\right]
  \end{eqnarray}
cf. Eqs.~\eref{partpro2} and \eref{ordsplit}. The dyadic matrix $\widetilde{A}$ has again a dual matrix
  \begin{eqnarray}\label{susysplit2}
   \widetilde{B}=\left[\begin{array}{cc} 0 & 0 \\ \sqrt{\widetilde{L}}V^\dagger &  0 \end{array}\right]\left[\begin{array}{cc} 0 & V\sqrt{\widetilde{L}}\\  0 & 0 \end{array}\right]=\left[\begin{array}{cc} 0 &  0 \\ 0 & \sqrt{\widetilde{L}}V^\dagger V\sqrt{\widetilde{L}} \end{array}\right],
  \end{eqnarray}
cf. Eq.~\eref{duality-new}. Interchanging the integrals over $\Omega$ and $V$ in Eq.~\eref{partchi3} we arrive at the following integral transform of $P$,
\begin{eqnarray}\label{transform}
  \Psi(\widetilde{A})&=&\int d[\Omega] P(\Omega^\dagger\Omega)\exp[-\imath\str \Omega\Omega^\dagger \widetilde{A}],
\end{eqnarray}
which plays the role of the Fourier-Laplace transform~\eref{Fourier-Laplace} in the case of Dyson's threefold way. Now the invariance of $\Omega$ under multiplication from the left with unitary supermatrices enters, implying
\begin{eqnarray}\label{invariancepsi}
  \Psi(\widetilde{U}\widetilde{A}\widetilde{U}^{-1})&=&\Psi(\widetilde{A}),\quad\forall\ \widetilde{U}\in\U^{(\beta)}(n+\widetilde{\gamma}k_2|\gamma k_2).
\end{eqnarray}
Hence, the following identity is true
\begin{eqnarray}\label{dualityiden}
  \Psi(\widetilde{A})&=&\Psi(\widetilde{B}),
\end{eqnarray}
connecting the ordinary matrix space with the superspace. This identity is remarkable, as it relates both spaces with one and the same superfunction $\Psi$. We notice that the supermatrices $\widetilde{A}$ and $\widetilde{B}$ are of the same size corresponding to the first and second splitting of Eq.~\eref{Chisplit2}, respectively, while their non-zero blocks are not.

The duality relation~\eref{dualityiden} can be plugged into the partition function which reads
\begin{eqnarray}
  \fl Z_\chi(\kappa)&=&(-1)^{\gamma (n+\nu)(k_2-k_1)}\sdet^{-\nu/\widetilde{\gamma}}\kappa\label{partchi4}\\
 \fl&&\times\frac{\int d[V]\int d[\Omega] P(\Omega^\dagger\Omega)\exp[\imath\str V^\dagger V \widetilde{L}\kappa^2-\imath\str \Omega\Omega^\dagger \widetilde{B}]}{\imath^{\gamma n(k_2-k_1)}\sdet^{-n/\widetilde{\gamma}} \widetilde{L}\int d[V]\exp[-\str V^\dagger V]\int d[\widehat{\Omega}] \exp[-\tr\widehat{\Omega}^\dagger\widehat{\Omega}]}.\nonumber
\end{eqnarray}
Due to convergence of the integrals we can again not easily switch the integration of $\Omega$ and $V$ unless the boson-boson block of $\widetilde{L}$ is proportional to the identity. However this problem can be circumvented as it was discussed in subsection~\ref{sec:brief} by introducing an auxiliary Hermitean supermatrix. We skip this here because it is exactly the same procedure explained in subsection~\ref{sec:brief}. Hence we end up with the partition function
\begin{eqnarray}\label{partchi5}
  \fl Z_\chi(\kappa)&=&(-1)^{\gamma (n+\nu)(k_2-k_1)}\sdet^{-\nu/\widetilde{\gamma}}\kappa\frac{\int d[\Omega] P(\Omega^\dagger\Omega)\sdet^{-n/\widetilde{\gamma}}(\Omega'\Omega^{'\,\dagger}-\kappa^2)}{\int d[\widehat{\Omega}] \exp[-\tr\widehat{\Omega}^\dagger\widehat{\Omega}]},
\end{eqnarray}
which is one of the main results of this section. We emphasize a few things about this formula. The supermatrices $\Omega'$ in the numerator and $\widehat{\Omega}$ in the denominator have different sizes, see the splittings~\eref{Chisplit2}. Moreover the index $\nu$ can take negative values as well since we have not at all used an assumption like $WW^\dagger$ is smaller than $W^\dagger W$. Equation~\eref{partchi5} can be slightly modified such that the supermatrix $\kappa$ can be easily assumed to be non-diagonal, e.g. in QCD you need a non-diagonal $\kappa$ to generate mixed pion condensates \cite{genmass}, or we can think of a symmetry breaking term in the determinant of Eq.~\eref{partchi2} which may happen by circumventing the problem of a two-sided correlated Wishart ensemble as it appears for modelling spatial-time correlation matrices~\cite{stcorr1,stcorr2,stcorr3}, see subsection~\ref{sec:normdep}.

The superdeterminant in Eq.~\eref{partchi5} only depends on the the product $\Omega'\Omega^{'\,\dagger}$. Therefore the integral over $W'$ defines a new probability distribution $\widehat{Q}$ on the superspace $\Gl^{(\beta)}(\widetilde{\gamma} k_1|\gamma k_2;n+\nu)$, i.e.
\begin{eqnarray}\label{projection2}
  \hspace*{-1cm}\widehat{Q}(\Omega^{\prime\,\dagger}\Omega')=\frac{\int d[W'] P(\Omega^\dagger\Omega)}{\int d[\widehat{\Omega}] \exp[-\tr\widehat{\Omega}^\dagger\widehat{\Omega}]}=\frac{\int d[W'] P(W^{\prime\,\dagger}W'+\Omega^{\prime\,\dagger}\Omega')}{\int d[\widehat{\Omega}] \exp[-\tr\widehat{\Omega}^\dagger\widehat{\Omega}]}.
\end{eqnarray}
Notice that there is one crucial disadvantage of this projection formula to the one of Dyson's threefold way, cf. Eq.~\eref{projection1}. The superfunction $\widehat{Q}$ is still a function depending on a matrix $\Omega^{\prime\,\dagger}\Omega'$ with ordinary dimensions. It is easy to get rid of this flaw if the original probability density $P$ is also invariant under right multiplication of $W$. Such a restriction becomes a problem for two-sided correlated Wishart matrices. For one-sided correlated Wishart matrix ensemble we can circumvent this problem, see subsection~\ref{sec:normdep}.

\subsection{Rotation invariant probability densities}
\label{sec:projsusy}

In this subsection we further simplify the projection formula by assuming that the probability density $P$ is rotation invariant, i.e.
\begin{eqnarray}\label{isotropy2}
 \fl P(W^\dagger W)=P(\widetilde{U}W^\dagger W\widetilde{U}^{-1}),\quad \forall\ \widetilde{U}\in \mathrm{U}^{(\beta)}(n+\nu)\ {\rm and}\ W\in\Gl^{(\beta)}(n;n+\nu).
\end{eqnarray}
Then this invariance is obviously true by  replacing $W\to\Omega$, too. Therefore there is certainly a supersymmtric extension of $P$ denoted by $\widetilde{P}$ with
\begin{eqnarray}\label{extension}
 P(W^{\prime\,\dagger}W'+\Omega^{\prime\,\dagger}\Omega')=\widetilde{P}\left(\left[\begin{array}{cc} W'W^{\prime\,\dagger} &  W'\Omega^{\prime\,\dagger} \\ \Omega'W^{\prime\,\dagger} & \Omega'\Omega^{\prime\,\dagger} \end{array}\right]\right).
\end{eqnarray}
The reason is that we can write $P$ in terms of matrix invariants like traces which is also a source of ambiguity when extending $P$ to superspace \cite{genHub2}.

For further calculations we assume $\nu\geq0$ which becomes important for convergence of some integrals. Because of the invariance under independent left and right multiplication of $W$ with unitary matrices this is not a restriction at all. One can simply choose $W$ such that it has the smaller dimension $n$ on its left side.

 Since the integral~\eref{projection2} is invariant under the transformation $\Omega^{\prime\,\dagger}\Omega'\to \widetilde{U}\Omega^{\prime\,\dagger}\Omega'\widetilde{U}^{-1}$ for all $\widetilde{U}\in\mathrm{U}^{(\beta)}(n)$, too,  we can define a probability density on superspace
\begin{eqnarray}\label{projection3}
  Q(\Omega'\Omega^{\prime\,\dagger})=\frac{\int d[W'] \widetilde{P}\left(\left[\begin{array}{cc} W'W^{\prime\,\dagger} &  W'\Omega^{\prime\,\dagger} \\ \Omega'W^{\prime\,\dagger} & \Omega'\Omega^{\prime\,\dagger} \end{array}\right]\right)}{\int d[\widehat{\Omega}] \exp[-\tr\widehat{\Omega}^\dagger\widehat{\Omega}]}.
\end{eqnarray}
The crucial difference of Eqs.~\eref{projection2} and \eref{projection3} is that $Q$ in contrast to $\widehat{Q}$ depends on  a $(\gamma\widetilde{\gamma}k_1|\gamma\widetilde{\gamma}k_2)\times(\gamma\widetilde{\gamma}k_1|\gamma\widetilde{\gamma}k_2)$ supermatrix. Thus, there is a chance to get rid of a number of integration variables which scales with $n$. This is quite important when taking the limit of large matrices as it is the case when deriving the universal behavior of the spectrum of $H_\chi$.

The aim is to express the integral~\eref{projection3} in terms of the combination $\Omega'\Omega^{\prime\,\dagger}$ and some integration variables. For this purpose we introduce Dirac $\delta$-functions for the blocks depending on $W'$,
\begin{eqnarray}\label{susycalc1}
  \fl Q(\Omega'\Omega^{\prime\,\dagger})&\propto&\int d[W']\int d[H_1]\int d[H_2] \int d[W_1] \int d[W_2] \widetilde{P}\left(\left[\begin{array}{cc}  H_1 &  W_1\\ W_1^\dagger & \Omega'\Omega^{\prime\,\dagger} \end{array}\right]\right)\\
  \fl&&\times\exp\left[\tr(H_1-W'W^{\prime\,\dagger})(\imath H_2 +\eins_{\gamma(n+\widetilde{\gamma}(k_2-k_1))})\right]\nonumber\\
  \fl&&\times\exp\left[\imath\tr(W_1-W'\Omega^{\prime\,\dagger} )W_2^\dagger+\imath\tr W_2(W_1^\dagger-\Omega'W^{\prime\,\dagger})\right].\nonumber
\end{eqnarray}
We drop the normalization constant right now and introduce it later on by fixing it with the Gaussian case. The matrices are drawn from $H_1,H_2\in\Herm^{(\beta)}(n+\widetilde{\gamma}(k_2-k_1))$ and $W_1^\dagger,W_2^\dagger\in\Gl^{(\beta)}(\widetilde{\gamma}k_1|\gamma k_2;n+\widetilde{\gamma}(k_2-k_1))$. Recall the definition of the cosets and the splitting of $\Omega$ in Eqs.~\eref{cosetdef}, \eref{Chienssusy} and \eref{Chisplit2}, respectively. The shift in $H_2$ guarantees the convergence of the integral over $W'$ which is the first one we perform yielding
\begin{eqnarray}\label{susycalc2}
  \fl Q(\Omega'\Omega^{\prime\,\dagger})&\propto&\lim\limits_{\delta\to0}\int d[H_1]\int d[H_2] \int d[W_1] \int d[W_2] \widetilde{P}\left(\left[\begin{array}{cc}  H_1 &  W_1\\ W_1^\dagger & \Omega'\Omega^{\prime\,\dagger} \end{array}\right]\right)\\
  \fl&&\times\exp\left[\tr H_1(\imath H_2 +\eins_{\gamma(n+\widetilde{\gamma}(k_2-k_1))})+\imath\tr W_1W_2^\dagger+\imath\tr W_2W_1^\dagger\right]\nonumber\\
  \fl&&\times\exp\left[-\tr(\imath H_2 +\eins_{\gamma(n+\widetilde{\gamma}(k_2-k_1))})^{-1}W_2(\Omega'\Omega^{\prime\,\dagger}+\delta\eins_{\gamma\widetilde{\gamma}k_1|\gamma\widetilde{\gamma}k_2})W_2^\dagger \right]\nonumber\\
  \fl&&\times{\det}^{-(n+\nu)/\widetilde{\gamma}}(\imath H_2 +\eins_{\gamma(n+\widetilde{\gamma}(k_2-k_1))}).\nonumber
\end{eqnarray}
The variable $\delta$ is a regularization guaranteeing us the convergence of the integrals since $\Omega'\Omega^{\prime\,\dagger}$ is not invertible if it contains a fermion-fermion block, i.e. $k_2\neq0$. We rescale $W_1\to W_1\sqrt{\Omega'\Omega^{\prime\,\dagger}+\delta\eins_{\gamma\widetilde{\gamma}k_1|\gamma\widetilde{\gamma}k_2}}$ and $W_2\to W_2/\sqrt{\Omega'\Omega^{\prime\,\dagger}+\delta\eins_{\gamma\widetilde{\gamma}k_1|\gamma\widetilde{\gamma}k_2}}$. The Jacobian of the transformation $W_1$ and $W_2$ cancel out and the limit of the regulator $\delta\to0$ can be made exact. The next integral we perform is over $W_2$ and we find
\begin{eqnarray}\label{susycalc3}
  \fl Q(\Omega'\Omega^{\prime\,\dagger})&\propto&\int d[H_1]\int d[H_2] \int d[W_1] \widetilde{P}\left(\left[\begin{array}{cc}  H_1 &  W_1\sqrt{\Omega'\Omega^{\prime\,\dagger}} \\ \sqrt{\Omega'\Omega^{\prime\,\dagger}}W_1^\dagger & \Omega'\Omega^{\prime\,\dagger} \end{array}\right]\right)\\
  \fl&&\times\exp\left[\tr (H_1-W_1W_1^\dagger)(\imath H_2 +\eins_{\gamma(n+\widetilde{\gamma}(k_2-k_1))})\right]\nonumber\\
  \fl&&\times{\det}^{-(n+\nu)/\widetilde{\gamma}+(k_1-k_2)}(\imath H_2 +\eins_{\gamma(n+\widetilde{\gamma}(k_2-k_1))}).\nonumber
\end{eqnarray}
We notice that $\widetilde{P}$ depends on invariants only. Hence in an explicit representation of $\widetilde{P}$ we do not encounter the ill-defined matrix $\sqrt{\Omega'\Omega^{\prime\,\dagger}} $ but only the supermatrix $\Omega'\Omega^{\prime\,\dagger}$. The remaining integral over $H_2$ is an ordinary Ingham-Siegel integral~\cite{Ingham,Siegel}. Shifting $H_1\to H_1+W_1W_1^\dagger$ the Ingham-Siegel integral tells us that $H_1$ has to be positive definite and yields a determinant of $H_1$ to the power $\nu/\widetilde{\gamma}+(\gamma-\widetilde{\gamma})/2$ (exactly here we need $\nu\geq0$). The positivity constraint of $H_1$ is quite often hard to handle such that we replace $H_1$ by a rectangular matrix $\widehat{W}_1\in\Gl^{(\beta)}(n+\widetilde{\gamma}(k_2-k_1);n+\nu+\widetilde{\gamma}(k_2-k_1))$. Finally, we arrive at the main result of this section and the projection formula for rotation invariant chiral ensembles,
\begin{eqnarray}\label{projection4}
  \fl Q(\Omega'\Omega^{\prime\,\dagger})&=&C\int d[\widehat{W}_1]\int d[W_1]  \widetilde{P}\left(\left[\begin{array}{cc}  \widehat{W}_1\widehat{W}_1^\dagger+W_1W_1^\dagger &  W_1\sqrt{\Omega'\Omega^{\prime\,\dagger}} \\ \sqrt{\Omega'\Omega^{\prime\,\dagger}}W_1^\dagger & \Omega'\Omega^{\prime\,\dagger} \end{array}\right]\right)
\end{eqnarray}
with the normalization constant
\begin{eqnarray}\label{norm}
\fl C=\frac{\int d[W']\exp[-\tr W'W^{\prime\,\dagger}]}{\int d[\widehat{W}_1]\exp[-\tr \widehat{W}_1\widehat{W}_1^{\dagger}]\int d[W_1]\exp[-\tr W_1W_1^{\dagger}]\int d[\widehat{\Omega}] \exp[-\tr\widehat{\Omega}^\dagger\widehat{\Omega}]}.
\end{eqnarray}
The reason for fixing the normalization with Gaussian weights lies in the universality of the projection formula~\eref{projection4}. The projection formula is true for almost all ensembles depending on invariants of the rectangular matrix $W$. Due to this broad applicability Eq.~\eref{projection4} is a powerful tool.  In Sec.~\ref{sec:examples}, we will present some examples, often encountered in different fields of random matrix theory.

Additionally one can apply the superbosonization formula to the partition function
\begin{eqnarray}\label{partchi5b}
  \fl Z_\chi(\kappa)&=&(-1)^{\gamma (n+\nu)(k_2-k_1)}\sdet^{-\nu/\widetilde{\gamma}}\kappa\int d[\Omega'] Q(\Omega'\Omega^{\prime\,\dagger})\sdet^{-n/\widetilde{\gamma}}(\Omega'\Omega^{'\,\dagger}-\kappa^2),
\end{eqnarray}
which is justified since the whole integral depends on the dyadic supermatrix $\Omega'\Omega^{\prime\,\dagger}$. Thus we replace $\Omega'\Omega^{\prime\,\dagger}$ by the supermatrix $\widehat{U}\in\Herm_{\odot}^{(\beta)}(\widetilde{\gamma}k_1|\gamma k_2)$ which has the same structure as the supermatrix $U$ in the original approach of the superbosonization formula~\eref{partition2}. The partition function reads
\begin{eqnarray}
  Z_\chi(\kappa)&=&\frac{(-1)^{\gamma (n+\nu)(k_2-k_1)}\int d[\Omega'] \exp[-\str \Omega'\Omega^{\prime\,\dagger}]}{\int d\mu(\widehat{U}) \exp[-\str \widehat{U}]\sdet^{(n+\nu)/\widetilde{\gamma}}\widehat{U}}\sdet^{-\nu/\widetilde{\gamma}}\kappa\nonumber\\
  &&\times\int d\mu(\widehat{U}) Q(\widehat{U})\sdet^{-n/\widetilde{\gamma}}(\widehat{U}-\kappa^2)\sdet^{(n+\nu)/\widetilde{\gamma}}\widehat{U}\label{partchi6}
\end{eqnarray}
with the superfunction
\begin{eqnarray}\label{projection5}
   \hspace*{-1cm}Q(\widehat{U})&=&C\int d[\widehat{W}_1]\int d[W_1]  \widetilde{P}\left(\left[\begin{array}{cc}  \widehat{W}_1\widehat{W}_1^\dagger+W_1W_1^\dagger &  W_1\sqrt{\widehat{U}} \\ \sqrt{\widehat{U}}W_1^\dagger & \widehat{U} \end{array}\right]\right).
\end{eqnarray}
Importantly, one should not confuse the superfunction $\widetilde{\Phi}$ of Eq.~\eref{partition2} with the superfunction $Q$, we mention the different terms in the integrands. The prefactor in Eq.~\eref{partchi6} is the global normalization constant resulting from the superbosonization formula and strongly depends on the normalization of the Haar-measure $d\mu(\widehat{U})$ of the supersymmetric coset $\Herm_{\odot}^{(\beta)}(\widetilde{\gamma}k_1|\gamma k_2)$. 

\section{Some examples}
\label{sec:examples}

We apply the projection formula~\eref{projection5} to four non-trivial examples to illustrate how our approach works. Especially it becomes clear what the advantages of the projection formula~\eref{projection5} are in comparison to the standard approaches with the generalized Hubbard-Stratonovich transformation~\cite{genHub1,genHub2} and the superbosonization formula~\cite{supform1,supform2}.

In particular we discuss norm-dependent ensembles and correlated Wishart ensembles in subsection~\ref{sec:normdep}, Lorentz-like (Cauchy) ensembles in subsection~\ref{sec:Lorentz}, the three unquenched chiral Gaussian ensembles in subsection~\ref{sec:unquenched}, and a probability density with a quartic potential in subsection~\ref{quartic}. The norm-dependent ensembles serve as a check since they can readily be calculated with the previous variants of the supersymmetry method. With help of the correlated Wishart ensembles we show that the projection formula can easily be extended to include a symmetry breaking constant term in the determinants, cf. Eq.~\eref{partordchi2}. The Lorentz-like (Cauchy) weight is another standard probability density as the Gaussian weight. It has a particular property namely it exhibits heavy tails and thus not all moments exist. For the unquenched chiral Gaussian ensemble we derive an alternative representation of the chiral Lagrangian, see Refs!
 .~\cite{RMTQCD1,RMTQCD2,RMTQCDbook} for the common representation. In this representation the physical mesons are split off from the artificial ones which result from introducing source terms to generate the desired observables. With help of the quartic potential we want to show that one can also study non-trivial potentials via the projection formula~\eref{projection5}.

\subsection{Norm-dependent ensembles and correlated Wishart ensembles}
\label{sec:normdep}

The first class of ensembles we want to look at are the norm-dependent chiral ensembles \cite{Mehta,normdep}, i.e.
\begin{equation}
P(W^\dagger W) = p(\tr W^\dagger W)
\label{eq:normdep}
\end{equation}
with an integrable function $p$. A particular choice is a fixed trace ensemble, namely $p(\tr W^\dagger W)\propto\delta(\tr W^\dagger W-cn)$ with a constant $c>0$. Such an ensemble naturally appears when modelling lattice QCD \cite{fixedtrace1}.  The lattice QCD Dirac operator is build up of unitary matrices and fulfills a fixed-trace condition. However one can readily show that this condition has only a minor effect on the microscopic regime of the Dirac spectrum and is completely suppressed in the exact limit \cite{fixedtrace1}. The choice $p\propto\delta(\tr W^\dagger W-cn)$ only enhances the $1/n$ correction. Also in quantum information it plays an important role~\cite{fixedtrace2} since the density operator is normalized.

The corresponding superfunction of the probability density $P$ for an arbitrary $p$ can be simply read off from the projection formula~\eref{projection5} and is up to a constant
\begin{equation}
Q(\widehat{U}) \propto \int_{0}^\infty dr p(r^2+\str \widehat U)r^{\beta(n+\widetilde{\gamma}(k_2-k_1))(n+\nu)}.
\label{eq:susynormdep}
\end{equation}
The exponent of the integration variable $r$ is the difference of the number of commuting real variables and anti-commuting Grassmann variables in the rectangular matrices $W_1$ and $\widehat{W}_1$. Those matrices are of dimension $(\gamma\widetilde{\gamma}k_1|\gamma\widetilde{\gamma}k_2)\times(\gamma n+\gamma\widetilde{\gamma}(k_2-k_1))$ and $(\gamma n+\gamma\widetilde{\gamma}(k_2-k_1))\times(\gamma (n+\nu)+\gamma\widetilde{\gamma}(k_2-k_1))$, respectively, and fulfil certain symmetries similar to Eq.~\eref{symmetries}.

A natural representative of a norm-dependent  ensemble is the Gaussian one, i.e. $p(\tr W^\dagger W)\propto\exp[-n\tr W^\dagger W]$. Then the integral over $r$ factorizes in Eq.~\eref{eq:susynormdep}. This apparently yields again a Gaussian
\begin{equation}
Q(\Omega'\Omega'^\dagger) \propto\exp\left(-\frac{n}{\widetilde{\gamma}}\mathrm{str}\Omega'\Omega'^\dagger\right)
\label{eq:GaussverteilungSuper1}
\end{equation}
in terms of the dyadic supermatrix $\Omega'\Omega'^\dagger$ and reads in terms of the supermatrix $\widehat{U}\in\Herm_{\odot}^{(\beta)}(\widetilde{\gamma}k_1|\gamma k_2)$
\begin{equation}
Q(\widehat{U}) \propto\exp\left(-\frac{n}{\widetilde{\gamma}}\mathrm{str}\widehat{U}\right).
\label{eq:GaussverteilungSuper2}
\end{equation}
For a Gaussian weight this result is not surprising but it serves as a simple check for the projection formula~\eref{projection5}. When plugging Eq.~\eref{eq:GaussverteilungSuper2} into the partition function~\eref{partchi6}, we arrive at
\begin{eqnarray}
  \fl Z_\chi(\kappa)&\propto&\sdet^{-\nu/\widetilde{\gamma}}\kappa\int d\mu(\widehat{U}) \exp\left(-n\,\mathrm{str}\widehat{U}\right)\sdet^{-n/\widetilde{\gamma}}(\widehat{U}-\kappa^2)\sdet^{(n+\nu)/\widetilde{\gamma}}\widehat{U}.\label{partgaus1}
\end{eqnarray}

The microscopic limit ($n\to\infty$ while $\nu$ and $n\kappa$ fixed) connects chiral random matrix theory with QCD~\cite{RMTQCDbook} and is obtained from our expression by rescaling $\widehat{U}\to-\imath\kappa\widehat{U}$. After taking the limit $n\to\infty$ we find the well-known chiral Lagrangian \cite{RMTQCDbook}
\begin{eqnarray}
 Z_\chi(\kappa)&\overset{n\gg1}{\propto}&\int d\mu(\widehat{U}) \exp\left(\imath \frac{n}{\widetilde{\gamma}}\mathrm{str}\,\kappa(\widehat{U}+U^{-1})\right)\sdet^{\nu/\widetilde{\gamma}}\widehat{U}.\label{partgaus2}
\end{eqnarray}
Surprisingly, we had not to take any saddlepoint approximation with our approach which is usually the case in the other approaches of the supersymmetry method~\cite{RMTQCD1,RMTQCD2}. The reason is that the projection formula already mapped the ordinary space to the correct coset describing the mesons of the chiral Lagrangian in QCD.

Another application of norm-dependent ensembles are correlated Wishart matrices with a non-Gaussian weight. In Sec.~\ref{sec:Zwei} we claimed that we can also study one-sided correlated Wishart ensembles with arbitrary weight. Those ensembles appear in many situations where one encounters time series analysis like in finance~\cite{finance1,finance2}, telecommunication~\cite{telecom1}, etc.  Thus we consider the following partition function
\begin{eqnarray}\label{partexa1.a}
\fl Z_\chi(\kappa)&=&(-1)^{\gamma (n+\nu)(k_2-k_1)}\sdet^{-\nu/\widetilde{\gamma}}\kappa\\
 \fl&&\times\int d[W] p(\tr W^\dagger C^{-1} W)\sdet^{-1/(\gamma\widetilde{\gamma})}(WW^\dagger\otimes\eins_{\gamma\widetilde{\gamma}k_1|\gamma\widetilde{\gamma}k_2}-\eins_{\gamma  n}\otimes\kappa^2),\nonumber
\end{eqnarray}
where the function $p$ is as before arbitrary and $C$ is an empirical correlation matrix and thus positive definite. In the first step we rescale $W\to \sqrt{C} W$ and have
\begin{eqnarray}\label{partexa1.b}
\fl Z_\chi(\kappa)&=&(-1)^{\gamma (n+\nu)(k_2-k_1)}\sdet^{-\nu/\widetilde{\gamma}}\kappa {\det}^{(n+\nu)/\widetilde{\gamma}+(k_2-k_1)} C\\
 \fl&&\times\int d[W] p(\tr W^\dagger W)\sdet^{-1/(\gamma\widetilde{\gamma})}(WW^\dagger\otimes\eins_{\gamma\widetilde{\gamma}k_1|\gamma\widetilde{\gamma}k_2}-C^{-1}\otimes\kappa^2).\nonumber
\end{eqnarray}
In the second step we apply the projection formula~\eref{projection5} in combination with a slightly modified version of Eq.~\eref{partchi6} and find
\begin{eqnarray}\label{partexa1.c}
  \fl Z_\chi(\kappa)&\propto&\sdet^{-\nu/\widetilde{\gamma}}\kappa{\det}^{(n+\nu)/\widetilde{\gamma}+(k_2-k_1)} C\\
  \fl&&\times\int d\mu(\widehat{U}) Q(\widehat{U})\sdet^{-1/(\gamma\widetilde{\gamma})}(\eins_{\gamma n}\otimes\widehat{U}-C^{-1}\otimes\kappa^2)\sdet^{(n+\nu)/\widetilde{\gamma}}\widehat{U}.\nonumber
\end{eqnarray}
The superfunction $Q$ is the one from the onefold integral~\eref{eq:susynormdep}. In the case of a Gaussian weight  the one-point correlation function was already studied with help of supersymmetry for $\beta=1,2$, see Refs.~\cite{Recher1,Recher2}. Equation~\eref{partexa1.c} is an alternative compact representation of this partition function.

\subsection{Lorentz (Cauchy)-like ensembles}
\label{sec:Lorentz}

Another kind of probability density serving as a `standard candle'  in statistical physics is the Lorentz weight. In contrast to the Gaussian weight, almost all moments of the matrix $W$ do not exist for the Lorentzian.  In random matrix theory one introduces this weight with a constant $\Gamma\in\mathds{R}_+$ determining the width of the distribution and an exponent $\mu \in \mathds{N}$  indicating how rapid the tails fall off, i.e. the Lorentzian ensemble is given by
\begin{equation}
P(W^\dagger W) \propto{\det}^{-\mu} \left(\Gamma^2 \mathds{1}_{n+\nu} + W^\dagger W\right).
\label{eq:lorentzverteilung}
\end{equation}
The exponent $\mu$ has to be large enough to guarantee the normalizability of the probability density. This ensemble is also known as Cauchy ensemble \cite{Cauchyens1,Cauchyens2}. Of particular interest is its heavy-tailed behavior which has not been studied in such detail as the exponential cut-off from ensembles with polynomial potentials. Importantly, one can expect that the universal results may break down. Recent works on heavy tails of random matrices are Refs.~\cite{heavytail1,heavytail2,heavytail3} and references therein.

Again we are interested in the supersymmtric analogue of $P$ which is given via the projection formula~\eref{projection5},
\begin{eqnarray}
   \fl Q(\widehat{U})&\propto&\int d[\widehat{W}_1]\int d[W_1]  {\sdet}^{-\mu} \left(\Gamma^2 \mathds{1}_{\gamma n+\gamma\widetilde{\gamma}k_2|\gamma\widetilde{\gamma}k_2} + \left[\begin{array}{cc}  \widehat{W}_1\widehat{W}_1^\dagger+W_1W_1^\dagger &  W_1\sqrt{\widehat{U}} \\ \sqrt{\widehat{U}}W_1^\dagger & \widehat{U} \end{array}\right]\right)\nonumber\\
   \fl&=&{\sdet}^{-\mu} \left(\Gamma^2 \mathds{1}_{\gamma\widetilde{\gamma}k_1|\gamma\widetilde{\gamma}k_2} +  \widehat{U} \right)\int d[\widehat{W}_1]\int d[W_1]\label{eq:lorentzverteilung2}\\
   \fl&&\times  {\sdet}^{-\mu} \left(\Gamma^2 \mathds{1}_{\gamma n+\gamma\widetilde{\gamma}(k_2-k_1)} + \widehat{W}_1\widehat{W}_1^\dagger+\Gamma^2 W_1(\Gamma^2 \mathds{1}_{\gamma\widetilde{\gamma}k_1|\gamma\widetilde{\gamma}k_2} +  \widehat{U} )^{-1}W_1^\dagger \right).\nonumber
\end{eqnarray}
In the second line we pulled out the lower right block of the superdeterminant. Here, we once more observe that one can often calculate with the superdeterminant as it would be a determinant, see Refs.~\cite{Berezin}. After rescaling $W_1\to W_1(\Gamma^2 \mathds{1}_{\gamma\widetilde{\gamma}k_1|\gamma\widetilde{\gamma}k_2} +  \widehat{U} )^{1/2}$ the integrals over $\widehat{W}_1$ and $W_1$ factorize and yield a constant. The projection formula leads to the superfunction (up to a normalization constant)
\begin{eqnarray}
   Q(\widehat{U})&\propto&{\sdet}^{n/\widetilde{\gamma}+(k_2-k_1)-\mu} \left(\Gamma^2 \mathds{1}_{\gamma\widetilde{\gamma}k_1|\gamma\widetilde{\gamma}k_2} +  \widehat{U} \right)\label{eq:lorentzverteilung3}.
\end{eqnarray}
Thus the counterpart of the Lorentzian weight~\eref{eq:lorentzverteilung} is also Lorentzian in superspace. Only the exponent changes. Notice that the fermion-fermion block of $\widehat{U}$ is a compact integral such that we do  not have any problems of convergence if $n/\widetilde{\gamma}+(k_2-k_1)-\mu\leq0$. The exponent $\mu$ has only to be large enough such that the corresponding partition function,
\begin{eqnarray}\label{partexa2.a}
  \fl Z_\chi(\kappa)&\propto&\sdet^{-\nu/\widetilde{\gamma}}\kappa\int d\mu(\widehat{U}) {\sdet}^{n/\widetilde{\gamma}+(k_2-k_1)-\mu} \left(\Gamma^2 \mathds{1}_{\gamma\widetilde{\gamma}k_1|\gamma\widetilde{\gamma}k_2} +  \widehat{U} \right)\\
  \fl&&\times\sdet^{-n/\widetilde{\gamma}}(\widehat{U}-\kappa^2)\sdet^{(n+\nu)/\widetilde{\gamma}}\widehat{U},\nonumber
\end{eqnarray}
exists, namely it has to be larger than $\mu>(n+\nu)/\widetilde{\gamma}$ for this integral. To guarantee the integral of the partition function in ordinary space the exponent has to fulfill $\mu>(n+\nu)/\widetilde{\gamma}+k_2-k_1$. Therefore one has only to take  $\mu>(n+\nu)/\widetilde{\gamma}+\max\{0,k_2-k_1\}$ to guarantee the convergence of both integrals. 

Interestingly, from Eq.~\eref{partexa2.a} immediately follows that in the microscopic limit $n\to\infty$ ($\nu$, $n\Gamma^2$ and $n\kappa$ fixed) for $\mu = n/\widetilde{\gamma}+\widetilde{\mu}$ with $\widetilde{\mu}$ fixed we do not find the universal result~\eref{partgaus2}. We already expected that something may change, i.e. the partition function becomes
\begin{eqnarray}
  \fl Z_\chi(\kappa)&\propto&\int d\mu(\widehat{U}) {\sdet}^{(k_2-k_1)-\widetilde{\mu}} \left(n\Gamma^2 \mathds{1}_{\gamma\widetilde{\gamma}k_1|\gamma\widetilde{\gamma}k_2} + n\kappa \widehat{U} \right)\sdet^{\nu/\widetilde{\gamma}}\widehat{U}\exp\left[\frac{n}{\widetilde{\gamma}}\str\kappa\widehat{U}^{-1}\right].\nonumber\\
  \fl&&\label{partexa2.b}
\end{eqnarray}
 However one can find the universal result at the hard edge of the spectrum, as the microscopic limit is also known, if $\widetilde{\mu}/n$ and $\Gamma^2$ is fixed instead.

\subsection{Unquenched chiral Gaussian ensemble}
\label{sec:unquenched}

The unquenched partition function is in QCD a statistical weight where additionally to the gauge action we have an interaction with fermionic quarks \cite{RMTQCDbook}. They are equivalent with additional characteristic polynomials in the numerator in the partition function. Hence, the random matrix model is
\begin{equation}
\fl P(WW^\dagger) =\frac{\exp\left(-n\tr W^\dagger W/\widetilde{\gamma}\right)\prod\limits_{j=1}^{N_{\rm f}}\mathrm{det}(W^\dagger W+m_j^2\eins_{\gamma(n+\nu)})}{\int d[W] \exp\left(-n\tr W^\dagger W/\widetilde{\gamma}\right)\prod\limits_{j=1}^{N_{\rm f}}\mathrm{det}(W^\dagger W+m_j^2\eins_{\gamma(n+\nu)})} 
\label{eq:unquenched}
\end{equation}
with the quark masses $m=\diag(m_1\eins_{\gamma\widetilde{\gamma}},\ldots, m_{N_{\rm f}}\eins_{\gamma\widetilde{\gamma}})$ of the $N_{\rm f}$ flavors. This time we explicitly wrote the normalization constant, since it is mass dependent and is, thus, quite essential.

The partition function~\eref{partordchi2} with the probability density~\eref{eq:unquenched}, i.e. the partially quenched partition function
\begin{eqnarray}\label{partquenched}
 \fl Z_\chi(\kappa,m)&=&\frac{(-1)^{\gamma (n+\nu)(k_2-k_1)}\sdet^{-\nu/\widetilde{\gamma}}\kappa}{\int d[W] \exp\left(-n\tr W^\dagger W/\widetilde{\gamma}\right)\prod\limits_{j=1}^{N_{\rm f}}\mathrm{det}(W^\dagger W+m_j^2\eins_{\gamma(n+\nu)})}\\
 \fl&&\times\int d[W] \exp\left(-\frac{n}{\widetilde{\gamma}}\tr W^\dagger W\right)\prod\limits_{j=1}^{N_{\rm f}}\mathrm{det}(W^\dagger W+m_j^2\eins_{\gamma(n+\nu)})\nonumber\\
 \fl&&\times\sdet^{-1/(\gamma\widetilde{\gamma})}(WW^\dagger\otimes\eins_{\gamma\widetilde{\gamma}k_1|\gamma\widetilde{\gamma}k_2}-\eins_{\gamma  n}\otimes\kappa^2),\nonumber
\end{eqnarray}
 can be dealt with in two different ways. Either the additional determinants and the determinants generating the correlation functions are computed on equal footing or one can consider the additional determinants as part of the probability density $P$. We decide for the latter choice since we aim at a separation of the physical quarks from the artificial ones which are also known as valence quarks.
 
In \ref{app1} we calculate the partially quenched partition function at finite $n$. It is a double integral over an ordinary matrix $U_{\pi}\in\Herm^{(\beta)}_\odot(0|\gamma N_{\rm f})=\U^{(4/\beta)}(\gamma N_{\rm f})$ and the supermatrix $\widehat{U}\in\Herm_{\odot}^{(\beta)}(\widetilde{\gamma}k_1|\gamma k_2)$,
\begin{eqnarray}
  \fl Z_\chi(\kappa,m)&\propto&\sdet^{-\nu/\widetilde{\gamma}}\kappa\biggl[\int d\mu(\widehat{U})\int d\mu(U_{\pi}) \exp\left(-\frac{n}{\widetilde{\gamma}} (\str\widehat{U}-\tr U_{\pi})\right)\nonumber\\
  \fl&&\times\sdet^{\nu/\widetilde{\gamma}}\widehat{U}\sdet^{-n/\widetilde{\gamma}}(\eins_{\gamma\widetilde{\gamma}k_1|\gamma\widetilde{\gamma}k_2}-\kappa^2\widehat{U}^{-1})\nonumber\\
  \fl&&\times{\det}^{\nu/\widetilde{\gamma}}U_{\pi}{\det}^{(n+\nu)/\widetilde{\gamma}+k_2-k_1}(\eins_{N_{\rm f}}+ m^2U_{\pi}^{-1})\nonumber\\
\fl&&\times \sdet^{1/(\gamma\widetilde{\gamma})}\left(\widehat{U}\otimes \eins_{\gamma\widetilde{\gamma}N_{\rm f}}+ \eins_{\gamma\widetilde{\gamma}k_1|\gamma\widetilde{\gamma}k_2}\otimes(U_{\pi}+ m^2)\right)\biggl]\biggl/\label{partque1}\\
\fl&&\biggl[\int d\mu(U_{\pi})\exp\left(\frac{n}{\widetilde{\gamma}}\tr U_{\pi}\right)\mathrm{det}^{(n+\nu)/\widetilde{\gamma}}( U_{\pi}+m^2){\det}^{-n/\widetilde{\gamma}}U_{\pi} \biggl].\nonumber
\end{eqnarray}
We take the microscopic limit $n\to\infty$ with $n\kappa$ and $nm$ fixed. The partially quenched  partition function becomes
\begin{eqnarray}
  \fl Z_\chi(\kappa,m)&\propto&\biggl[\int d\mu(\widehat{U})\int d\mu(U_{\pi})\sdet^{\nu/\widetilde{\gamma}}\widehat{U}{\det}^{\nu/\widetilde{\gamma}}U_{\pi}\nonumber\\
  \fl&&\times \exp\left(\frac{n}{\widetilde{\gamma}} \tr m(U_{\pi}+U^{-1}_\pi)-\frac{n}{\widetilde{\gamma}} \str\kappa(\widehat{U}-\widehat{U}^{-1})\right)\nonumber\\
\fl&&\times \sdet^{1/(\gamma\widetilde{\gamma})}\left(n\kappa\widehat{U}\otimes \eins_{\gamma\widetilde{\gamma}N_{\rm f}}+ \eins_{\gamma\widetilde{\gamma}k_1|\gamma\widetilde{\gamma}k_2}\otimes nmU_{\pi}\right)\biggl]\biggl/\label{partque2}\\
\fl&&\biggl[\int d\mu(U_{\pi})\exp\left(\frac{n}{\widetilde{\gamma}}\tr m(U_{\pi}+U_{\pi}^{-1})\right){\det}^{\nu/\widetilde{\gamma}}U_{\pi} \biggl].\nonumber
\end{eqnarray}
This partition function has to agree with the well-known results for the three chiral ensembles, see Refs.~\cite{RMTQCD1,RMTQCD2,RMTQCDbook}. It is equal to Eq.~\eref{partgaus2} when the variables $\kappa$ also comprise the quark masses $m$. For $\beta=2$ this can be readily checked due to the knowledge of the Harish-Chandra-Itzykson-Zuber integral \cite{Harish,ItzZub}. In the real and quaternion case this is not as easy since the corresponding group integrals are not known.

What is the benefit of the representation~\eref{partque2} of the partially quenched partition function? The physical quarks are completely separated from the auxiliary particles, i.e. the chiral Lagrangian for the physical mesons can be read off
\begin{eqnarray}\label{chiLag}
  \fl\mathcal{L}(U_{\pi},\kappa,m)&=&\frac{n}{\widetilde{\gamma}} \tr m(U_{\pi}+U^{-1}_\pi)+{\rm ln}\biggl[\int d\mu(\widehat{U})\exp\left(-\frac{n}{\widetilde{\gamma}} \str\kappa(\widehat{U}-\widehat{U}^{-1})\right)\sdet^{\nu/\widetilde{\gamma}}\widehat{U}\nonumber\\
  \fl&&\times\sdet^{1/(\gamma\widetilde{\gamma})}\left(n\kappa\widehat{U}\otimes \eins_{\gamma\widetilde{\gamma}N_{\rm f}}+ \eins_{\gamma\widetilde{\gamma}k_1|\gamma\widetilde{\gamma}k_2}\otimes nmU_{\pi}\right)\biggl].
\end{eqnarray}
The first part of the Lagrangian is the leading order of the unquenched partition function with $N_{\rm f}$ flavors \cite{RMTQCDbook}. The second term is the operator corresponding to the generating function for some observables like the level density. Therefore we could split the observable from the physical system, $U_\pi$, in the chiral Lagrangian with help of the projection formula. Since random matrix theory only describes the Goldstone bosons with zero momentum a good question is if one can achieve such a splitting~\eref{chiLag} for the kinetic modes, too.

\subsection{Probability density with quartic potential}
\label{quartic}

In the last example we want to consider the probability density with quartic potential
\begin{eqnarray}\label{quarticprob}
 P(WW^\dagger)&\propto&\exp[-\alpha\tr(WW^\dagger)^2-\widehat{\alpha}\tr WW^\dagger],
\end{eqnarray}
$\alpha>0$ and $\widehat{\alpha}\in\mathbb{R}$. This probability density is the standard one for the analysis of multicritical behavior \cite{multicrit1,multicrit2,multicrit3,multicrit4}. Depending on the relation of the two constants $\alpha$ and $\widehat{\alpha}$ the macroscopic level density of $WW^\dagger$ can exhibit a one-cut or two-cut solution which also influences the universality on the local scale of the mean level density where the two cuts are merging to one. We are aiming at a supersymmetric representation of the partition function with the probability density~\eref{quarticprob}.

The superfunction $Q$ corresponding to the probability density~\eref{quarticprob} is via the projection formula~\eref{projection5}
\begin{eqnarray}
   \fl Q(\widehat{U})&\propto&\int d[\widehat{W}_1]\int d[W_1]  \exp\left[-\alpha\left(\tr(\widehat{W}_1\widehat{W}_1^\dagger+W_1W_1^\dagger)^2+\tr W_1\widehat{U}W_1^\dagger+\str\widehat{U}^2\right)\right]\nonumber\\
   \fl&&\times\exp\left[-\widehat{\alpha}\left(\tr\widehat{W}_1\widehat{W}_1^\dagger+\tr W_1W_1^\dagger +\str\widehat{U}\right)\right].\label{quartsusy1}
\end{eqnarray}
The quartic term $\tr(\widehat{W}_1\widehat{W}_1^\dagger+W_1W_1^\dagger)^2$ can be traced back to a quadratic structure by introducing a Gaussian over an auxiliary matrix $H\in\Herm^{(\beta)}(n+\widetilde{\gamma}(k_2-k_1))$. Then the integrals over $\widehat{W}_1$ and $W_1$ are purely Gaussian and can be performed without any problem, leading to
\begin{eqnarray}
   \fl Q(\widehat{U})&\propto&\exp\left[-\alpha\str\widehat{U}^2-\widehat{\alpha}\str\widehat{U}\right]\int d[H]  \exp\left[-\frac{1}{4\alpha}\tr(H-\imath(\widehat{\alpha}-1)\eins_{\gamma n+\gamma\widetilde{\gamma}(k_2-k_1)})^2\right]\nonumber\\
   \fl&&\hspace*{-0cm}\times{\det}^{-(n+\nu)/\gamma+k_1-k_2}(\imath H-\eins_{\gamma n+\gamma\widetilde{\gamma}(k_2-k_1)})\nonumber\\
   \fl&&\hspace*{-0cm}\times\sdet^{-1/(\gamma\widetilde{\gamma})}(\imath H\otimes\eins_{\gamma\widetilde{\gamma} k_1|\gamma\widetilde{\gamma} k_2}-\eins_{\gamma n+\gamma\widetilde{\gamma}(k_2-k_1)}\otimes(\alpha\widehat{U}+\eins_{\gamma\widetilde{\gamma} k_1|\gamma\widetilde{\gamma} k_2})).\nonumber\\
   \fl&&\label{quartsusy2}
\end{eqnarray}
The determinant results from the integral over $\widehat{W}_1\in\Gl^{(\beta)}(n+\widetilde{\gamma}(k_2-k_1);n+\nu+\widetilde{\gamma}(k_2-k_1))$ while the superdeterminant results from the integral over $W_1^\dagger\in\Gl^{(\beta)}(\widetilde{\gamma} k_1|\gamma k_2;n+\widetilde{\gamma}(k_2-k_1))$. We recall the definition of the cosets in Eqs.~\eref{cosetdef}, \eref{Chiens}, and \eref{Chienssusy}. The shift in the Gaussian of the auxiliary ordinary matrix $H$ also guarantees the convergence of the integrals over $\widehat{W}_1$ and $W_1$ for negative $\widehat{\alpha}$.

For $\beta=2$ the integral~\eref{quartsusy2} can be further simplified via various techniques in random matrix theory \cite{Mehta,detstruc1,detstruc2,detstruc3}. In one of these techniques \cite{Mehta,detstruc3} one constructs the orthogonal polynomials of the weight $g(E)=\exp[-(E-\imath(\widehat{\alpha}-1))^2/(4\alpha)]/(\imath E-1)^{n+\nu+k_1-k_2}$. Then one obtains a quotient of two determinants of $ \max\{k_1,k_2\}\times\max\{k_1,k_2\}$ matrices where the determinant in the numerator depends on the orthogonal polynomials and their Cauchy transform with respect to the weight $g(E)$ whose arguments are the eigenvalues of the supermatrix $(\alpha\widehat{U}+\eins_{\gamma\widetilde{\gamma} k_1|\gamma\widetilde{\gamma} k_2})$. The determinant in the denominator is the square root of the Berezinian (Jacobian in superanalysis) resulting from a diagonalization of the supermatrix $\widehat{U}$ \cite{detstruc2}. See Refs.~\cite{Mehta,detstruc3} and references therein for an intro!
 duction in the application of orthogonal polynomials.

For $\beta=1,4$ the situation is not as simple. Though the ordinary matrix $H$ is decoupled from the supermatrix $\widehat{U}$ and no unknown group integrals make the calculation insurmountable, the square root of the superdeterminant hinders the application of orthogonal polynomial theory. The obvious way out of this dilemma is the expansion of the integral~\eref{quartsusy2} in the matrix $\widehat{U}$. Then one can calculate each of the expansion coefficients. Since $H$ and $\widehat{U}$ are decoupled such an expansion is trivial. The non-trivial task is to perform the integral over $H$ to find the coefficients. We emphasize that such an expansion is finite if $k_1=0$ because the superdeterminant becomes a determinant in the numerator and, thus, a polynomial in $\widehat{U}$.

What is the benefit of $Q$, see Eq.~\eref{quartsusy2}, in particular when there is no explicit, simple expression? The advantage of the result~\eref{quartsusy2} with the corresponding partition function in superspace is revealed when considering the correlated situation, meaning that we destroy the invariance of $W$ under the multiplication from the right (or left) with unitary matrices by an external correlation matrix $C$. In contrast to the partition function in ordinary space with the probability weight~\eref{quarticprob} we do not encounter large group integrals (if $k_1$ and $k_2$ are small) when diagonalizing $H$. The resulting partition function is Eq.~\eref{partexa1.c} where we replace the norm-dependent superfunction by the superfunction~\eref{quartsusy2}. Particularly the calculation of the level density is capable in this way for all three Dyson indices, see Refs.~\cite{Recher1,Recher2} for the Gaussian ensemble.

\section{Summary and Conclusions}
\label{sec:summary}

We presented a new variant of the supersymmetry method which directly
relates the probability density in ordinary space with the one in
superspace via a projection formula. Thereby we briefly rederived this
formula, see Eq.~\eref{projection1}, for the ensembles originally
included in Dyson's threefold way \cite{Cartan1}, namely real
symmetric, Hermitean, and Hermitean self-dual matrices, which was
first done in Ref.~\cite{Doktor}. In a second step we extended the
idea behind such a projection formula to the three chiral
ensembles. Hereby we found a formula for ensembles whose invariance of
the rectangular matrices under multiplication from the right (or left)
is broken, see Eq.~\eref{projection2}. This formula is quite
convenient for those situations when introducing empirical correlation
matrices on both sides of the rectangular random matrix as it is the
case in spacial-temporal correlations \cite{stcorr1,stcorr2,stcorr3}.

The result~\eref{projection2} is not as compact as the further
simplified formula~\eref{projection5} which is only possible if we
ensure the invariance of the probability density under left and right
multiplication of the rectangular random matrix with unitary
matrices. The supersymmetric integral in the partition function is
over one of the three coset integrals depending on the Dyson index
$\beta$ which already play a crucial role in the standard approach
with the superbosonization
formula~\cite{supform1,supform2}. Nevertheless one should not confuse
our approach with the one in Refs.~\cite{supform1,supform2}.

The projection formula~\eref{projection1} for the three non-chiral
ensembles agrees with the result of the generalized
Hubbard-Stratonovich transformation~\cite{genHub1,genHub2} in the
integration domain as well as in the form of the integrand. This is
not the case for the chiral ensembles where the projection formula
shares the integration domain with the original superbosonization
formula~\cite{supform1,supform2} while the integrand is of a
completely different form and resembles more the one of the
generalized Hubbard-Stratonovich
transformation~\cite{genHub1,genHub2}. Therefore the projection
formulas~\eref{projection2} and \eref{projection5} for chiral
ensembles represent an alternative approach to the standard
supersymmetry methods in random matrix theory.

We applied the projection formula~\eref{projection5} to the relatively
simple example of norm-dependent ensembles and found a quite compact
and explicit dependence of the probability density in superspace on
the one in ordinary space which reduces to a onefold
integral~\eref{eq:susynormdep}. For the Gaussian case we recovered the
well known chiral Lagrangian of QCD \cite{RMTQCD1,RMTQCD2,RMTQCDbook}
in the microscopic limit, see Eq.~\eref{partgaus2}. Furthermore we
showed how to generalize the projection formula in the case of
one-sided correlated random matrices, see Eq.~\eref{partexa1.c}. This
underlines that the projection formula~\eref{projection5} is not at
all restricted to rotation invariant (`isotropic'~\cite{isotropy})
ensembles but can also cover a simple, but also the most popular kind
of symmetry breaking.

Another ensemble to which we applied the projection
formula~\eref{projection5} is of Lorentz (Cauchy) type. Surprisingly
not only the Gaussian weight is form-invariant under mapping the
probability density in ordinary space to one in superspace but also
the Lorentz weight. Only the exponent of the determinant changes and
has to be taken care of. With help of the representation in superspace
we showed that depending on the exponent of the determinant the
Lorentzian shows universal behavior in the microscopic limit or
not. Hence the projection formula~\eref{projection5} provides a new
tool to investigate universality issues in chiral random matrix
theory, as well.

Moreover, we considered the standard application of chiral random
matrix theory to QCD. With help of the projection
formula~\eref{projection5} we split the chiral Lagrangian of the
partially quenched theory in QCD into two parts, see
Eq.~\eref{chiLag}. One part consists of the lowest order of the
unquenched theory in the physical mesons (the pions for two flavors)
which is the well-known linear term in the quark masses
\cite{RMTQCD1,RMTQCD2,RMTQCDbook}. We refer to the expansion scheme of
the microscopic limit (the limit of large space-time volume,
$V\to\infty$, with fixed rescaled quark masses, $Vm={\rm const.}$),
see Refs.~\cite{RMTQCDbook}, which is one kind of a low energy
expansion. The other part represents the interaction with the source
terms which are artificially introduced to generate the
observables. This is some kind of a natural splitting into the
physical system and the measurement. It would be quite interesting if
such a splitting is also applicable to the kinetic modes of the mesons
which are not included in the lowest order description by random
matrix theory. Maybe chiral perturbation theory can shed light to
this.

In a fourth example we considered a probability density with a quartic
potential emphasizing that the projection formula~\eref{projection5}
can also deal with more complicated situations. We derived a
representation of the probability density in superspace which is still
an integral over a Hermitian matrix $H$, see
Eq.~\eref{quartsusy2}. However the coupling of the ordinary matrix $H$
with the supermatrix $\widehat{U}$ is in an invariant way, meaning
that $H$ and $\widehat{U}$ are independently invariant under unitary
transformations. In the case of the Dyson index $\beta=2$ this allows
to apply the machinery of orthogonal
polynomials~\cite{Mehta,detstruc3} and other
techniques~\cite{detstruc1,detstruc2} (whereby Ref.~\cite{detstruc1}
is not limited to $\beta = 2$) to calculate an
explicit expression of the probability density $Q$, see Eq.~\eref{quartsusy2}. For an
elaborate presentation of the calculation methods we refer to
Ref.~\cite{RMTbook}. In the other two cases $\beta=1,4$ the situation
is not as simple. Nevertheless we showed how to circumvent unknown
group integrals via the projection formula~\eref{projection5} in the
supersymmetry method if one considers one-sided correlated rectangular
random matrices drawn from an ensemble with a quartic potential. For
the Gaussian case two of the authors already applied the supersymmetry
method to correlated Wishart ensemble and derived a compact expression
for the level density, see Refs.~\cite{Recher1,Recher2}. The
projection formula~\eref{projection5} opens a way to perform this
calculation for other probability densities as well.

\section*{Acknowledgements}

We acknowledge support from the Deutsche Forschungsgemeinschaft within
Sonderforschungsbereich Transregio 12 ``Symmetries and Universality in
Mesoscopic Systems'' (VK, TG) and partial financial support from the
Alexander von Humboldt foundation (MK). Moreover we thank Gernot
Akemann and Jacobus J. M. Verbaarschot for fruitful discussions.

\appendix

\section{The derivation of  Eq.~\eref{partque1}}\label{app1}

 Considering the partially quenched partition function~\eref{partquenched} with the probability density~\eref{eq:unquenched}, the integral that has to be performed via the projection formula~\eref{projection5} is
\begin{eqnarray}
\fl Q(\widehat{U})& \propto&\frac{1}{\int d[W] \exp\left(-n\tr W^\dagger W/\widetilde{\gamma}\right)\prod\limits_{j=1}^{N_{\rm f}}\mathrm{det}( WW^\dagger+m_j^2\eins_{\gamma n})}\nonumber\\
\fl&&\times \int d[\widehat{W}_1]\int d[W_1]  \exp\left(-\frac{n}{\widetilde{\gamma}} [\tr \widehat{W}_1\widehat{W}_1^\dagger+\tr W_1W_1^\dagger+\str\widehat{U}]\right)\nonumber\\
\fl&&\times\prod\limits_{j=1}^{N_{\rm f}}\sdet\left(\left[\begin{array}{cc}  \widehat{W}_1\widehat{W}_1^\dagger+W_1W_1^\dagger &  W_1\sqrt{\widehat{U}} \\ \sqrt{\widehat{U}}W_1^\dagger & \widehat{U} \end{array}\right]+m_j^2\eins_{\gamma n+\gamma\widetilde{\gamma}k_2|\gamma\widetilde{\gamma}k_2}\right)\nonumber\\
\fl& =&\frac{\exp\left[-n\str\widehat{U}/\widetilde{\gamma}\right]\prod\limits_{j=1}^{N_{\rm f}}\sdet\left(\widehat{U}+m_j^2\eins_{\gamma\widetilde{\gamma}k_1|\gamma\widetilde{\gamma}k_2}\right)}{\int d[W] \exp\left(-n\tr W^\dagger W/\widetilde{\gamma}\right)\prod\limits_{j=1}^{N_{\rm f}}\mathrm{det}( WW^\dagger+m_j^2\eins_{\gamma n})}\label{eq:unquenched2}\\
\fl&&\times\int d[\widehat{W}_1]\int d[W_1]  \exp\left(-\frac{n}{\widetilde{\gamma}} [\tr \widehat{W}_1\widehat{W}_1^\dagger+\tr W_1W_1^\dagger]\right)\nonumber\\
\fl&&\times\prod\limits_{j=1}^{N_{\rm f}}\det\left( \widehat{W}_1\widehat{W}_1^\dagger+m_j^2W_1(\widehat{U}+m_j^2\eins_{\gamma\widetilde{\gamma}k_1|\gamma\widetilde{\gamma}k_2})^{-1}W_1^\dagger+m_j^2\eins_{\gamma n +\gamma\widetilde{\gamma}(k_2-k_1)}\right).\nonumber
\end{eqnarray}
In the second step we pushed out the block matrices $\widehat{U}+m_j^2\eins_{\gamma\widetilde{\gamma}k_1|\gamma\widetilde{\gamma}k_2}$ for each mass $m_j$. The product of determinants can be rewritten as
\begin{eqnarray}
\fl&&\det\left( \widehat{W}_1\widehat{W}_1^\dagger+m_j^2W_1(\widehat{U}+m_j^2\eins_{\gamma\widetilde{\gamma}k_1|\gamma\widetilde{\gamma}k_2})^{-1}W_1^\dagger+m_j^2\eins_{\gamma n +\gamma\widetilde{\gamma}(k_2-k_1)}\right)\label{detrewrite}\\
\fl&=&m_j^{-2\gamma\nu}\sdet^{-1}(\widehat{U}+m_j^2\eins_{\gamma\widetilde{\gamma}k_1|\gamma\widetilde{\gamma}k_2})\,\sdet\left(\widetilde{W}_1^\dagger\widetilde{W}_1+\widehat{U}'+m_j^2 \eins_{\gamma(n+\nu)+\gamma\widetilde{\gamma}k_2|\gamma\widetilde{\gamma}k_2}\right)\nonumber
\end{eqnarray}
with
\begin{eqnarray}\label{Wdef}
 \widetilde{W}_1=\left[\begin{array}{cc} \widehat{W}_1 & W_1 \end{array}\right]
\end{eqnarray}
such that $\widetilde{W}_1^\dagger\in\Gl^{(\beta)}(n+\nu+\widetilde{\gamma}k_2|\widetilde{\gamma}k_2;n+\widetilde{\gamma}(k_2-k_1))$ and with the supermatrix
\begin{eqnarray}\label{Mdef}
 \widehat{U}'=\left[\begin{array}{cc} 0 & 0 \\  0 & \widehat{U} \end{array}\right].
\end{eqnarray}
The superfunction $Q$ reads
\begin{eqnarray}
\fl Q(\widehat{U})& \propto&\frac{\exp\left[-n\str\widehat{U}/\widetilde{\gamma}\right]}{\int d[W] \exp\left(-n\tr W^\dagger W/\widetilde{\gamma}\right)\prod\limits_{j=1}^{N_{\rm f}}\mathrm{det}( W^\dagger W+m_j^2\eins_{\gamma (n+\nu)})}\int d[\widehat{W}_1]\label{eq:unquenched3}\\
\fl&&\hspace*{-1cm}\times \exp\left(-\frac{n}{\widetilde{\gamma}} \str \widetilde{W}_1^\dagger\widetilde{W}_1\right)\sdet^{1/(\gamma\widetilde{\gamma})}\left([\widetilde{W}_1^\dagger\widetilde{W}_1+\widehat{U}']\otimes \eins_{\gamma\widetilde{\gamma}N_{\rm f}}+ \eins_{\gamma(n+\nu)+\gamma\widetilde{\gamma}k_2|\gamma\widetilde{\gamma}k_2}\otimes m^2\right). \nonumber
\end{eqnarray}
The integral over the supermatrix $\widetilde{W}_1$ resembles the partition function~\eref{partexa1.b} with an external matrix $\widehat{U}'$. One can easily show that the projection formula~\eref{projection5} can be generalized to a partition function with rotation invariant probability density in superspace. Thus we apply the projection formula for norm-dependent ensembles, see Eq.~\eref{partexa1.c}, to replace the dyadic supermatrix $\widetilde{W}_1^\dagger\widetilde{W}_1$ with a $\gamma\widetilde{\gamma}N_{\rm f}\times\gamma\widetilde{\gamma}N_{\rm f}$ unitary matrix $U_{\pi}\in\Herm^{(\beta)}_\odot(0|\gamma N_{\rm f})=\U^{(4/\beta)}(\gamma N_{\rm f})$ in the second tensor space in the superdeterminant~\eref{eq:unquenched3}. We recall the definitions~\eref{eq:InvarianzGruppe} and \eref{superHerm}. The subscript ``$\pi$'' of the unitary matrix $U$ refers to physical mesons as they indeed agree with the mesons (Goldstone bosons) in the microscopic limit. For $N_{\rm f}=2$ !
 the mesons are the pions which are usually denoted by $\pi$.

Employing the projection formula~\eref{projection5} to the expression~\eref{eq:unquenched3} the superfunction $Q$ takes the form
\begin{eqnarray}
\fl Q(\widehat{U})& \propto&\frac{\exp\left[-n\str\widehat{U}/\widetilde{\gamma}\right]}{\int d\mu(U_{\pi})\exp\left(n\tr U_{\pi}/\widetilde{\gamma}\right)\mathrm{det}^{(n+\nu)/\widetilde{\gamma}}( U_{\pi}+m^2)\det^{-n/\widetilde{\gamma}}U_{\pi} }\nonumber\\
\fl&&\times\int d\mu(U_{\pi})\exp\left(-\frac{n}{\widetilde{\gamma}} \tr U_{\pi}\right){\det}^{-n/\widetilde{\gamma}-k_2+k_1}U_{\pi}\nonumber\\
\fl&&\times \sdet^{1/(\gamma\widetilde{\gamma})}\left(\widehat{U}'\otimes \eins_{\gamma\widetilde{\gamma}N_{\rm f}}+ \eins_{\gamma(n+\nu)+\gamma\widetilde{\gamma}k_2|\gamma\widetilde{\gamma}k_2}\otimes(U_{\pi}+ m^2)\right) \nonumber\\
\fl& =&\frac{\exp\left[-n\str\widehat{U}/\widetilde{\gamma}\right]}{\int d\mu(U_{\pi}) \exp\left(n\tr U_{\pi}/\widetilde{\gamma}\right)\mathrm{det}^{(n+\nu)/\widetilde{\gamma}}( U_{\pi}+m^2)\det^{-n/\widetilde{\gamma}}U_{\pi} }\nonumber\\
\fl&&\times\int d\mu(U_{\pi})\exp\left(\frac{n}{\widetilde{\gamma}} \tr U_{\pi}\right){\det}^{\nu/\widetilde{\gamma}}U_{\pi}{\det}^{(n+\nu)/\widetilde{\gamma}+k_2-k_1}(\eins_{N_{\rm f}}+ m^2U_{\pi}^{-1})\nonumber\\
\fl&&\times \sdet^{1/(\gamma\widetilde{\gamma})}\left(\widehat{U}\otimes \eins_{\gamma\widetilde{\gamma}N_{\rm f}}+ \eins_{\gamma\widetilde{\gamma}k_1|\gamma\widetilde{\gamma}k_2}\otimes(U_{\pi}+ m^2)\right). \label{eq:unquenched4}
\end{eqnarray}
We also replaced the integral in the denominator via the projection formula. The superfunction $Q$ can be plugged into the partition function~\eref{partquenched} and we find Eq.~\eref{partque1}. 

\section*{References}

\begin{thebibliography}{x}
\begin{footnotesize}
	\bibitem{Wishart} 		J. Wishart, Biometrika {\bf 20}, 32 (1928).
	
	\bibitem{timeseries1} 	C. Chatfield, {\it The Analysis of Time Series} (Chapman and Hall/CRC, Boca Raton, 6th ed., 2004).
    \bibitem{timeseries2}	M. M\"uller, G. Baier, A. Galka, U. Stephani, and H. Muhle, Phys. Rev. E {\bf 71}, 046116 (2005).
	\bibitem{timeseries3}	P. \v{S}eba, Phys. Rev. Lett. {\bf 91}, 198104 (2003).
	
	\bibitem{RMTbook}		G. Akemann, J. Baik, and P. Di Francesco (Eds.), {\it The Oxford Handbook of Random Matrix Theory} (Oxford University Press, 1st ed., Oxford, 2011).
	
	\bibitem{OrthogonalPol} Y.V. Fyodorov and G. Akemann, JETP Letters {\bf 77}, 438 (2003)
	\bibitem{toda} K. Splittorff and J.J.M. Verbaarschot, Phys.Rev.Lett. {\bf 90} (2003) 041601
	\bibitem{mapHamilton} C.A. Tracy and H. Widom, Comm. Math. Phys. {\bf 159}, 151 (1994)
	\bibitem{mapHamilton2} C.A. Tracy and H. Widom, Comm. Math. Phys. {\bf 161}, 289 (1994)
	
	\bibitem{timeseries4}	N. El Karoui, {\it Multivariate Statistics}, Chapter 28 in Ref.~\cite{RMTbook}.
	
	\bibitem{finance1}		L. Laloux, P. Cizeau, J. P. Bouchaud, and M. Potters, Phys. Rev. Lett. {\bf 83}, 1467 (1999).
	\bibitem{finance2}		J.-P. Bouchard and M. Potters, {\it Financial Applications}, Chapter 40 in Ref.~\cite{RMTbook}.
	
	\bibitem{Recher1}		C. Recher, M. Kieburg, and T. Guhr, Phys. Rev. Lett. {\bf 105}, 244101 (2010).
	\bibitem{Recher2}		C. Recher, M. Kieburg, T. Guhr, and M. R. Zirnbauer, J. Stat. Phys. {\bf 148}, 981 (2012).
	
   	\bibitem{RMTQCD1}	E. V. Shuryak and J. J. M. Verbaarschot, Nucl. Phys. A {\bf 560}, 306 (1993).
   	\bibitem{RMTQCD2}	J. J. M. Verbaarschot, Phys. Lett. B {\bf 329}, 351 (1994); Phys. Rev. Lett. {\bf 72}, 2531 (1994).
   
   	\bibitem{RMTQCDbook}	J. J. M. Verbaarschot, {\it Quantum Chromodynamics}, Chapter 32 in Ref.~\cite{RMTbook}.
   
   	\bibitem{RMTQCDdata1}	J. J. M. Verbaarschot and T. Wettig,  Ann. Rev. Nucl. Part. Sci. {\bf 50}, 343 (2000).
   	\bibitem{RMTQCDdata2} P. H. Damgaard, Nucl. Phys. Proc. Suppl. {\bf 106}, 29 (2002).
   
   	\bibitem{conmat1}			C. W. J. Beenakker, {\it Condensed Matter Physics}, Chapter 35 in Ref.~\cite{RMTbook}.
   
   	\bibitem{telecom1}		A. Tulino and S. Verd\'u, {\it Random Matrix Theory And Wireless Communications} (Now Publishers, Hanover, MA, 2004); {\it Information Theory}, Chapter 41 in Ref.~\cite{RMTbook}.
	\bibitem{telecom2}		G. Akemann, M. Kieburg, and L. Wei,  J. Phys. A {\bf 46}, 275205 (2013).
	\bibitem{telecom3}		G. Akemann, J. R. Ipsen, and M. Kieburg, Phys. Rev. E {\bf 88}, 052118 (2013).
  
   \bibitem{inftheo1}		S. N. Majumdar, {\it Extreme Eigenvalues of Wishart Matrices and Entangled Bipartite System}, Chapter 37 in Ref.~\cite{RMTbook}.
   
   \bibitem{univloc1}		E. Br\'{e}zin and A. Zee, Nucl. Phys. B {\bf 402}, 613 (1993); C. R. Acad. Sci. Paris {\bf t.317}, 735 (1993).
   \bibitem{univloc2}		G. Hackenbroich and H. A. Weidenm{\"u}ller, Phys. Rev. Lett. {\bf 74}, 4118 (1995).
   \bibitem{univloc3}		G. Akemann, P. H. Damgaard, U. Magnea, and S. Nishigaki, Nucl. Phys. B {\bf 487}, 721 (1997).
   \bibitem{univloc4}		A. Kuijlaars, {\it Universality}, Chapter 6 in Ref.~\cite{RMTbook}.
   
   \bibitem{SUSYGauss1}	K. B. Efetov,  Adv. in Phys. {\bf 32}, 53–127 (1983); \textit{Supersymmetry in disorder and chaos soliton fractals}, (Cambridge University Press, Cambridge, 1997). 
   \bibitem{SUSYGauss2}	J. J. M. Verbaarschot, H. A. Weidenm{\"u}ller, and M. R. Zirnbauer. Phys. Rep. {\bf 129}, 367 (1985).
   \bibitem{SUSYGauss3}	M. R. Zirnbauer, \textit{Supersymmetry Methods in Random Matrix Theory}, Encyclopedia of Mathematical Physics {\bf 5}, 151 (Eds. J.-P. Franoise, G.L. Naber, and S.T. Tsou, Elsevier, Oxford, 2006).
   \bibitem{SUSYGauss4}	T. Guhr: {\it Supersymmetry}, Chapter 7 in Ref.~\cite{RMTbook}.
   
   \bibitem{univglob1}		D. Voiculescu, Lecture Notes in Mathematics {\bf 1738}, 279 (Springer, 2000).
   \bibitem{univglob2}		R. Speicher, {\it Free Probability Theory}, Chapter 22 in Ref.~\cite{RMTbook}.
   
   \bibitem{isotropy}			Z. Burda, M. A. Nowak, and A. Swiech, Phys. Rev. E {\bf 86}, 061137 (2012).
   
   \bibitem{genHub1}		T. Guhr, J. Phys. A {\bf 39}, 13191 (2006).
   \bibitem{genHub2}		M. Kieburg, J. Gr\"onqvist, and T. Guhr, J. Phys. A {\bf 42}, 275205 (2009).
   
   \bibitem{supform1}		H.-J. Sommers, Acta Phys. Pol. B {\bf 38}, 1001 (2007).
   \bibitem{supform2}		P. Littelmann, H.-J. Sommers, and M. R. Zirnbauer, Commun. Math. Phys. {\bf 283}, 343 (2008).
   
   \bibitem{comp} M. Kieburg, H.J. Sommers und T. Guhr, \textit{Comparison of the superbosonization formula and the generalized Hubbard-Stratonovich transformation}, J. Phys. A: Math. Theor. 42, 275206, (2009).

	\bibitem{Cartan1}		F. J. Dyson, J. Math. Phys. {\bf 3}, 1199 (1962).
	\bibitem{Cartan2}		M. R. Zirnbauer, J. Math. Phys. {\bf 37}, 4986 (1996); {\it Symmetry Classes}, Chapter 3 in Ref.~\cite{RMTbook}.
   
   \bibitem{Doktor}			M. Kieburg, {\it Supersymmetry in Random Matrix Theory} (dissertation at the Universit\"at Duisburg-Essen, Germany, 2010), DOI: http://duepublico.uni-duisburg-essen.de/servlets/DerivateServlet/Derivate-24404/Doktorarbeit\_Mario\_Kieburg.pdf.
   
   \bibitem{Cauchy1}		F. Wegner, unpublished notes (1983).
   \bibitem{Cauchy2}		F. Constantinescu, J. Stat. Phys. {\bf 50}, 1167 (1988).
   \bibitem{Cauchy3}		F. Constantinescu and H. F. de Groote, J. Math. Phys. {\bf 30}, 981 (1989).
   \bibitem{Cauchy4}		M. Kieburg, H. Kohler, and T. Guhr, J. Math. Phys. {\bf 50}, 013528 (2009).
   
   \bibitem{genmass}		B. Klein, D. Toublan, and J. J. M. Verbaarschot, Phys. Rev. D {\bf 68}, 014009 (2003); Phys. Rev. D {\bf 72}, 015007, (2005).
   
   \bibitem{GUEtrans1}		T. Guhr, Ann. Phys. {\bf 250}, 145 (1996).
   \bibitem{GUEtrans2}		G. Akemann and T. Nagao, JHEP {\bf 10}, 060 (2011).
 
   \bibitem{Mehta}			M. L. Mehta, \textit{Random Matrices} (Academic Press Inc., New York, 3rd edition, 2004).
   
   \bibitem{finiteQCD1}	M. A. Stephanov, Phys. Lett. B {\bf 375}, 249 (1996); Phys. Rev. Lett. {\bf 76}, 4472 (1996); Nucl. Phys. Proc. Suppl. {\bf 53}, 469 (1997).
   \bibitem{finiteQCD2}	T. Guhr and T. Wettig, Nucl. Phys. B {\bf 506}, 589 (1997).
   \bibitem{finiteQCD3}	J. C. Osborn, Phys. Rev. Lett. {\bf 93}, 222001 (2004).
   
	\bibitem{VKG}				V. Kaymak, M. Kieburg, and T. Guhr, 2014, in preparation.   
	
	\bibitem{stcorr1}			A. Ribes, J.-M. Aza\"is, and S. Planton, Clim. Dyn. {\bf 35}, 391 (2010).
	\bibitem{stcorr2}			J. L. Toole, N. Eagle, and J.B. Plotkin, ACM Trans. Intell. Syst. Technol. {\bf 2}, 38 (2011).
	\bibitem{stcorr3}			M. Snarska, Acta Phys. Pol. A {\bf 121}, B-110 (2012).
   
   \bibitem{Ingham}			A. E. Ingham, Proc. Camb. Phil. Soc. {\bf 29},271 (1933).
   \bibitem{Siegel}			C. L. Siegel, Ann. Math. {\bf 36}, 527 (1935).
   
   \bibitem{normdep}		T. Guhr,  J. Phys. A {\bf 39}, 12327 (2006).
   
   \bibitem{fixedtrace1}	M. Kieburg, J. J. M. Verbaarschot, and S. Zafeiropoulos, arXiv:1405.0433 [hep-lat] (2014).
   \bibitem{fixedtrace2}	G. Akemann and P. Vivo, J. Stat. Mech. {\bf 05}, 05020 (2011).
   
   \bibitem{Cauchyens1}	N. S. Witte and P. J. Forrester, Nonlinearity {\bf 13}, 1965 (2000); Nagoya Math. J. {\bf 174}, 29 (2004).
   \bibitem{Cauchyens2}	J. Najnudel, A. Nikeghbali, and F. Rubin, J. Stat. Phys. {\bf 137}, 373 (2009).
   
   \bibitem{heavytail1}		S. Mendelson and G. Paouris, arXiv:1108.3886 [math.PR] (2011).
   \bibitem{heavytail2}		R. A. Davis, O. Pfaffel, and R. Stelzer, arXiv:1108.5464v2 [math.PR] (2012).
   \bibitem{heavytail3}		Z. Burda and J. Jurkiewicz, {\it Heavy-Tailed Random Matrices}, Chapter 13 in Ref.~\cite{RMTbook}.
   
   \bibitem{Berezin}			F. A. Berezin, {\it Introduction to Superanalysis} (D. Reidel Publishing Company, Dordrecht, 1st ed., 1987).
   
   \bibitem{Harish}			Harish-Chandra,  Am. J. Math. {\bf 80}, 241 (1958).
   \bibitem{ItzZub}			C. Itzykson und J.B. Zuber, J. Math. Phys. {\bf 21}, 411 (1980).
   
   \bibitem{multicrit1}		Y. Shimamune, Phys. Lett. B {\bf 108}, 407 (1982).
   \bibitem{multicrit2}		G. M. Cicuta, L. Molinari, and E. Montaldi, Mod. Phys. Lett. A {\bf 1}, 125 (1986); J. Phys. A {\bf 23}, L421 (1990).
   \bibitem{multicrit3}		L. Molinari, J. Phys. A {\bf 21}, 1 (1988).
   \bibitem{multicrit4}		M. Douglas, N. Seiberg, and S. Shenker, Phys. Lett. B {\bf 244}, 381 (1990).
   
   \bibitem{detstruc1}		A. Borodin and E. Strahov, Commun. Pure Appl. Math. {\bf 59}, 161 (2005).
   \bibitem{detstruc2}		M. Kieburg and T. Guhr, J. Phys. A {\bf 43}, 075201 (2010).
   \bibitem{detstruc3}		M. Adler, {\it Spectral Statistics of Orthogonal and Symplectic Ensembles}, Chapter 5 in Ref.~\cite{RMTbook}.
\end{footnotesize}
\end{thebibliography}

\end{document}